\documentclass[reprint,aps,nofootinbib]{revtex4-1}
\usepackage{graphicx}
\usepackage{float}
\usepackage{amsmath}
\usepackage{amssymb}
\usepackage{verbatim}
\usepackage{array}
\newcolumntype{P}[1]{>{\centering\arraybackslash}p{#1}}
\usepackage{tikz}
\usetikzlibrary{calc,matrix}
\usepackage{hyperref}
\usepackage[utf8]{inputenc}
\usepackage{xcolor}
\hypersetup{
    colorlinks,
    linkcolor={red!50!black},
    citecolor={blue!50!black},
    urlcolor={blue!80!black}
}
\usepackage[T1]{fontenc}
\usepackage{textcomp}
\usepackage{diagbox}
\usepackage{listings}
\usepackage{scalerel}
\newcommand{\tpm} {\text{T}\phi\text{M}}

\newcommand{\tb} {\bar{3}}

\newcommand{\xb} {\bar{6}}

\newcommand{\om} {\omega}
\newcommand{\ob} {\bar{\omega}}
\newcommand{\Ta} {\tau}
\newcommand{\Tb} {\bar{\tau}}
\newcommand{\sgm} {\Sigma(72\times3)}

\newcommand{\ceo} {\mathbf{C(\tau)_{\scaleto{\!\!\!~1}{4pt}}}}
\newcommand{\cet} {\mathbf{C(\tau)_{\scaleto{\!\!\!~2}{4pt}}}}
\newcommand{\cer} {\mathbf{C(\tau)_{\scaleto{\!\!\!~3}{4pt}}}}
\newcommand{\cef} {\mathbf{C(\tau)_{\scaleto{\!\!\!~4}{4pt}}}}
\newcommand{\cro} {\mathbf{C(\omega)_{\scaleto{\!\!\!~1}{4pt}}}}
\newcommand{\crt}{\mathbf{C(\omega)_{\scaleto{\!\!\!~2}{4pt}}}}
\newcommand{\crr} {\mathbf{C(\omega)_{\scaleto{\!\!\!~3}{4pt}}}}
\newcommand{\crf} {\mathbf{C(\omega)_{\scaleto{\!\!\!~4}{4pt}}}}
\newcommand{\crv} {\mathbf{C(\omega)_{\scaleto{\!\!\!~5}{4pt}}}}
\newcommand{\ctw} {\mathbf{C(\tau\omega)_{\scaleto{\!\!\!~1}{4pt}}}}
\newcommand{\A} {\mathbf{A}}
\newcommand{\B} {\mathbf{B}}
\newcommand{\Bm} {\mathbf{|B|}}
\newcommand{\Do} {\mathbf{D_{\scaleto{\!\!\!~1}{4pt}}}}
\newcommand{\Dt} {\mathbf{D_{\scaleto{\!\!\!~2}{4pt}}}}
\newcommand{\Eo} {\mathbf{E_{\scaleto{\!\!\!~1}{4pt}}}}
\newcommand{\Et} {\mathbf{E_{\scaleto{\!\!\!~2}{4pt}}}}
\newcommand{\Di} {\mathbf{D_{\scaleto{\!\!\!~i}{4pt}}}}
\newcommand{\Dj} {\mathbf{D_{\scaleto{\!\!\!~j}{4pt}}}}
\newcommand{\Ei} {\mathbf{E_{\scaleto{\!\!\!~i}{4pt}}}}
\newcommand{\Ej} {\mathbf{E_{\scaleto{\!\!\!~j}{4pt}}}}
\newcommand{\bt} {X_{24}}

\newcommand{\phia} {\acute{\phi}}
\newcommand{\phib} {\grave{\phi}}
\newcommand{\bind} {\Delta}
\usepackage[figuresright]{rotating}
\newcommand*{\p}{\hspace{-0.2mm}\turnbox{16}{$\,'$}\;}
\newcommand {\rtp}{\boldsymbol{3\p}}

\begin{document}
\preprint{APS/123-QED}

\title{Fully constrained mass matrix: Can symmetries alone determine the flavon vacuum alignments?}
\author{R.~Krishnan}
\email{krishnan.rama@saha.ac.in}
\homepage{\\https://orcid.org/0000-0002-0707-3267}
\affiliation{Saha Institute of Nuclear Physics, 1/AF Bidhannagar, Kolkata 700064, India}  
\date{\today}
\begin{abstract}
In the framework of the representation theory of finite groups, it was recently shown that a fully constrained complex-symmetric mass matrix can be conveniently mapped into a sextet of $\sgm$. In this paper, we introduce an additional flavor group $\bt$ in the model so that the vacuum alignment of the $\sgm$ sextet is determined not only by the symmetries of $\sgm$ but also by that of $\bt$. We define several flavons which transform as multiplets under $\sgm$ as well as $\bt$. The vacuum alignment of each of these flavons is obtained as a simultaneous invariant eigenstate of specific elements of the groups $\sgm$ and $\bt$; i.e.,~the vacuum alignment is fully determined by its residual symmetries. These flavons couple together uniquely resulting in the fully constrained sextet of $\sgm$. Through this work we propose a general formalism in which the flavor symmetry group ($G_f$) is obtained as the direct product, $G_f=G_r \times G_x$. Fermions transform nontrivially only under $G_r$ while they remain invariant under $G_x$. Flavons, on the other hand, transform nontrivially under both $G_r$ and $G_x$. The vacuum alignment of each flavon multiplet transforming irreducibly under $G_r \times G_x$ is uniquely identified by its corresponding residual symmetry (a subgroup of $G_r \times G_x$). Several such flavons couple together to form an effective multiple of $G_r$ which remains invariant under $G_x$. This effective multiplet couples to the fermions.
  
\end{abstract}
%
\maketitle

\section{Introduction}

More than two decades~\cite{Bilenky:2016pep} of experiments in neutrino oscillations have provided us with measurements of the neutrino mixing angles $\theta_{12}$, $\theta_{23}$, $\theta_{13}$ as well as the mass-squared differences, $\Delta m^2_{21}$, $\Delta m^2_{31}$~\cite{Giganti:2017fhf, Esteban:2016qun}. Yet, several features of neutrinos remain a mystery. Ordering of neutrino masses, $CP$ violation in neutrino sector, nature of neutrinos (Majorana or Dirac), and existence of sterile neutrinos are some of them. Parameters such as the light neutrino mass and the complex phases in the mixing matrix also need to be measured. Many of these questions are expected to be resolved by future experiments in the coming decades~\cite{Capozzi:2018ubv, deSalas:2017kay, Lattanzi:2017ubx, Kudenko:2017nti, Chatterjee:2017xkb, DiDomizio:2017rkc, Cao:2017drk}.

The initial measurements of large solar ($\theta_{12}$) and atmospheric ($\theta_{23}$) mixing angles stimulated the theoretical study of flavor symmetries in the neutrino sector based on discrete finite groups~\cite{Altarelli:2010gt, Grimus:2011mp, King:2013eh, Meloni:2017cig}. Tribimaximal mixing~\cite{Harrison:2002er} with $\theta_{12}=\sin^{-1}(1/\sqrt{3})$, $\theta_{23}=\pi/4$, and $\theta_{13}=0$ was widely used as a template for building models in the neutrino sector. With the measurement of the nonzero reactor  ($\theta_{13}$) mixing angle inconsistent with tribimaximal mixing, theorists have turned to alternative mixing schemes. A natural approach is to extend tribimaximal mixing with one or more free parameters~\cite{He:2011gb, Gupta:2011ct, Garg:2013xwa, Dev:2016bml, Ma:2007ku, Pakvasa:2008zz, King:2007pr, Plentinger:2005kx, Antusch:2013kna, Antusch:2013ti}. One such ansatz, called tri-phi-maximal mixing ($\tpm$)~\cite{Harrison:2002kp}, leads to a mixing matrix of the form
\begin{equation}
U_{\tpm}=\left(\begin{matrix}\sqrt{\frac{2}{3}}\cos \phi & \frac{1}{\sqrt{3}} & \sqrt{\frac{2}{3}}\sin \phi\\
-\frac{\cos \phi}{\sqrt{6}}-\frac{\sin \phi}{\sqrt{2}} & \frac{1}{\sqrt{3}} & \frac{\cos \phi}{\sqrt{2}}-\frac{\sin \phi}{\sqrt{6}}\\
-\frac{\cos \phi}{\sqrt{6}}+\frac{\sin \phi}{\sqrt{2}} & \frac{1}{\sqrt{3}} & -\frac{\cos \phi}{\sqrt{2}}-\frac{\sin \phi}{\sqrt{6}}
\end{matrix}\right).
\end{equation}
The angle $\phi$ parametrizes the nonzero reactor mixing angle. Like tribimaximal mixing, tri-phi-maximal mixing also has a trimaximal second column and is $CP$ conserving.

\begin{figure}[]
\begin{center}
\includegraphics[scale=0.7]{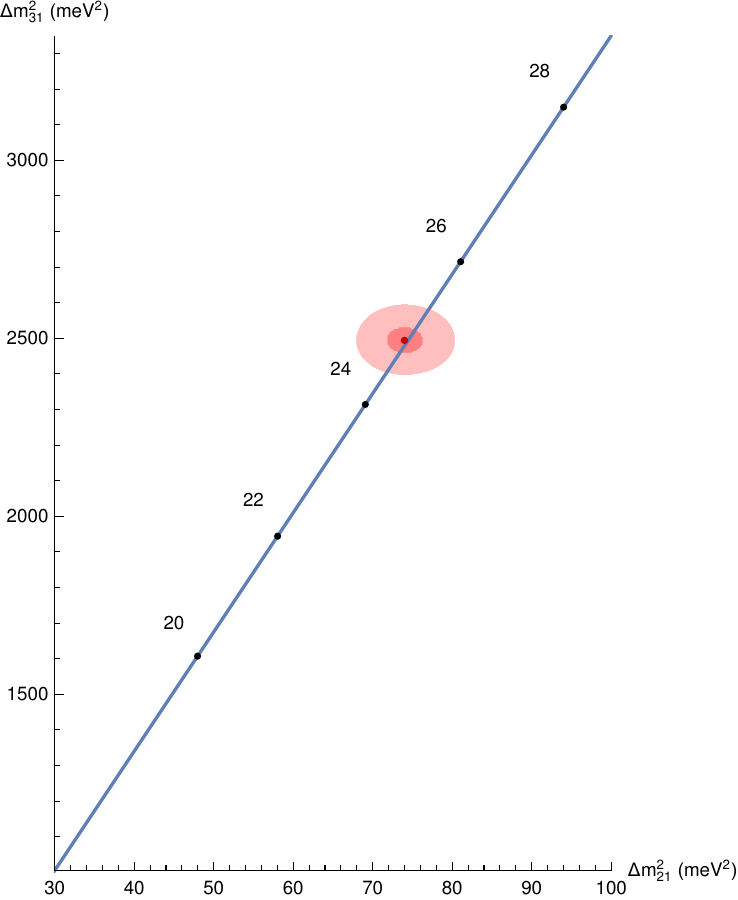}
\caption[$\Delta m_{31}^2$ vs $\Delta m_{21}^2$ plane]{The neutrino mass ratios Eq.~(\ref{eq:numass}) represented as a straight line in the $\Delta m_{31}^2$ vs $\Delta m_{21}^2$ plane. We have $\Delta m_{21}^2=m_1^2(\frac{1}{2}\tan^2\left(\frac{5\pi}{16}\right)-1)$ and $\Delta m_{31}^2=m_1^2(\tan^4\left(\frac{5\pi}{16}\right)-1)$ where $m_1$ is the lightest neutrino mass. The black dots denote $m_1$ in the units of milli-electron-Volt. The experimental range of $\Delta m_{21}^2$ and $\Delta m_{31}^2$ corresponds to the red dot and the shaded regions (the best fit value along with the $1\sigma$ and the $3\sigma$ errors).}
\label{fig:neutrinopredict2}
\end{center}
\end{figure}

An ansatz of neutrino Majorana mass matrices,
\begin{equation}\label{eq:tpmmat}
M_\text{Maj} \propto \left(\begin{matrix}i+\frac{1-i}{\sqrt{2}} & 0 & 1-\frac{1}{\sqrt{2}}\\
0 & 1 & 0\\
1-\frac{1}{\sqrt{2}} & 0 & -i+\frac{1+i}{\sqrt{2}}
\end{matrix}\right)\!\!,\,\, \propto \left(\begin{matrix}-i+\frac{1+i}{\sqrt{2}} & 0 & 1-\frac{1}{\sqrt{2}}\\
0 & 1 & 0\\
1-\frac{1}{\sqrt{2}} & 0 & i+\frac{1-i}{\sqrt{2}}
\end{matrix}\right),
\end{equation}
which leads to $\tpm$ with $\phi=\pm\pi/16$ was proposed~\cite{Krishnan:2012sb} shortly after the discovery of nonzero $\theta_{13}$ by the Daya Bay experiment. These matrices are fully constrained in the sense that they do not contain free parameters. In Ref.~\cite{Krishnan:2012sb}, the neutrino Dirac mass matrix was constructed to be proportional to the identity. As a result, the light neutrino masses obtained through the type-1 seesaw mechanism become proportional to the inverse of the eigenvalues of the Majorana mass matrices, Eqs.~(\ref{eq:tpmmat}). They are given by\footnote{Given as $\frac{\left(2+\sqrt{2}\right)}{1+\sqrt{2(2+\sqrt{2})}}:1:\frac{\left(2+\sqrt{2}\right)}{-1+\sqrt{2(2+\sqrt{2})}}$ in Ref.~\cite{Krishnan:2012sb}}
\begin{equation}\label{eq:numass}
m_1:m_2:m_3=\sqrt{2}\tan\left(\frac{3\pi}{16}\right):1:\sqrt{2}\tan\left(\frac{5\pi}{16}\right).
\end{equation} 
These are consistent with the measured neutrino mass-squared differences and also predict the experimentally undetermined light neutrino mass to be around $25$~meV. In Fig.~\ref{fig:neutrinopredict2}, we compare these ratios with the experimental mass-squared differences.

Recently~\cite{Krishnan:2018tja} it was shown that the group $\sgm$ can be used to model such fully constrained mass matrices. $\sgm$ can be obtained using four generators, namely $C$, $E$, $V$ and $X$~\cite{Grimus:2010ak}. For the three-dimensional representation, we have
\begin{equation}\label{eq:gen3}
\begin{split}
&C \equiv
\left(\begin{matrix}1 & 0 & 0\\
       0 & \om & 0\\
       0 & 0 & \ob
\end{matrix}\right), \quad \quad \quad \, \, \, E \equiv
\left(\begin{matrix}0 & 1 & 0\\
       0 & 0 & 1\\
       1 & 0 & 0
\end{matrix}\right),\\
&V\equiv
-\frac{i}{\sqrt{3}}\left(\begin{matrix}1 & 1 & 1\\
       1 & \om & \ob\\
       1 & \ob & \om
\end{matrix}\right), \quad X\equiv
-\frac{i}{\sqrt{3}}\left(\begin{matrix}1 & 1 & \ob\\
       1 & \om & \om\\       
       \om & 1 & \om
       \end{matrix}\right).
\end{split}
\end{equation}
The tensor product expansion of two triplets of this group is given by 
\begin{align}
\boldsymbol{3}\otimes\boldsymbol{3}&=\boldsymbol{6}\oplus\boldsymbol{\tb},\label{eq:tensor1}\\
\boldsymbol{\tb}\otimes\boldsymbol{\tb}&=\boldsymbol{\bar{6}}\oplus\boldsymbol{3}.\label{eq:tensor2}
\end{align}
$\sgm$ is the smallest group that produces a complex sextet from the tensor product of two identical triplets as shown in Eqs.~(\ref{eq:tensor1}) and (\ref{eq:tensor2}). Note that the triplets of the continuous group $SU(3)$ also have the same tensor product expansion. In Ref.~\cite{Krishnan:2018tja}, we assigned the right-handed neutrinos to be a conjugate triplet,
\begin{equation}
\nu_R=(\nu_{R1},\nu_{R2},\nu_{R3})^T \equiv \boldsymbol{\tb}.
\end{equation}
In the Majorana mass term, two of these conjugate triplets couple to produce a conjugate sextet,
\begin{equation}\label{eq:conjsextet}
\sum_{jk} S_{ijk} \nu_{Rj}\cdot\nu_{Rk} \equiv \boldsymbol{\xb}_i, 
\end{equation}
where $\nu_{Rj}\cdot\nu_{Rk}$ are the Lorentz invariant products of the right-handed neutrino Weyl spinors. $S_{ijk}$ are the familiar Clebsch-Gordan (C-G) coefficients for the symmetric tensor product of two triplets of $SU(3)$. We use the conventional basis where the nonzero coefficients are given by
\begin{equation}\label{eq:cg}
\begin{split}
S_{111}&= S_{222}= S_{333}=1, \\
S_{423}&=S_{432}= S_{531}= S_{513}=S_{612}=S_{621}=\frac{1}{\sqrt{2}}.
\end{split}
\end{equation}
We also introduced a flavon sextet, 
\begin{equation}\label{eq:xinu}
\xi=(\xi_1,\xi_2,\xi_3,\xi_4,\xi_5,\xi_6)^T \equiv \boldsymbol{6},
\end{equation}
which couples with the conjugate sextet, Eq.~(\ref{eq:conjsextet}), to produce the $\sgm$-invariant mass term, 
\begin{equation}\label{eq:Tnu}
\begin{split}
&\sum_{ijk}S_{ijk} \, \xi_i\,  \nu_{Rj}\cdot\nu_{Rk}\\
&\quad\quad= \left(\begin{matrix}\nu_{R1}\\
	\nu_{R2}\\
\nu_{R3}
\end{matrix}\right)^T \left(\begin{matrix}\xi_1 & \frac{1}{\sqrt{2}}\xi_6 & \frac{1}{\sqrt{2}}\xi_5\\
       \frac{1}{\sqrt{2}}\xi_6 &\xi_2 & \frac{1}{\sqrt{2}}\xi_4\\
       \frac{1}{\sqrt{2}}\xi_5 & \frac{1}{\sqrt{2}}\xi_4 &\xi_3
\end{matrix}\right)\cdot\left(\begin{matrix}\nu_{R1}\\
	\nu_{R2}\\
\nu_{R3}
\end{matrix}\right).
\end{split}
\end{equation}
The flavon sextet acquires a vacuum expectation value (VEV) through spontaneous symmetry breaking (SSB), and this VEV determines the structure of the mass matrix. Comparing Eq.~(\ref{eq:xinu}) with Eq.~(\ref{eq:Tnu}), it is clear that there is a one-to-one correspondence between the components of the sextet and the elements of the $3\times3$ complex-symmetric Majorana mass matrix. A specific VEV of the sextet fully constrains the mass matrix. The VEVs which correspond to the Majorana mass matrices, Eqs.~(\ref{eq:tpmmat}), are
\begin{align}
\langle\xi\rangle &\propto \left(i+\frac{1-i}{\sqrt{2}},1,-i+\frac{1+i}{\sqrt{2}},0, (\sqrt{2}-1), 0\right)^T,\label{eq:vevtpmp}\\
&\propto\left(-i+\frac{1+i}{\sqrt{2}},1,i+\frac{1-i}{\sqrt{2}},0, (\sqrt{2}-1), 0\right)^T.\label{eq:vevtpmm}
\end{align}
In Ref.~\cite{Krishnan:2018tja} we constructed flavon potentials that, through SSB, resulted in these VEVs, and this reproduced the mass matrices, Eqs.~(\ref{eq:tpmmat}). These mass matrices are diagonalized by $2\times2$ unitary matrices. The mixing matrix of the form $\tpm(\phi=\pm\frac{\pi}{16})$ is obtained as the product of a $3\times 3$ trimaximal contribution from the charged-lepton sector and the above-mentioned $2\times2$ contribution from the neutrino sector. The mixing angles extracted from $\tpm(\phi=\pm\frac{\pi}{16})$ are quite close to the experimental values. We used higher order corrections in the charged-lepton sector to account for the small discrepancy between the $\tpm(\phi=\pm\frac{\pi}{16})$ and the experimental values.

\section{Vacuum Alignment in Flavor Space}

In this section, we briefly review the salient features of model building using flavons. Even though the principles discussed are applicable for modeling mass matrices involving various fermions, here we study the Majorana mass matrix involving three families of right-handed neutrinos. The three neutrino states are assumed to form a triplet under a discrete flavor group, in general a subgroup of the continuous group, $U(3)$. To construct the Majorana mass terms, we calculate the tensor product expansion of two such triplets. This expansion gives rise to several neutrino-neutrino terms that transform as various multiplets under the flavor group. We theorize the existence of flavons that also transform as multiplets under the flavor group. The neutrino-neutrino multiplets and the corresponding flavon multiplets (conjugates) couple, leading to flavor group invariant mass terms. Through SSB, the flavons acquire VEVs. These VEVs and the coupling constants appearing along with the invariant mass terms constitute the Majorana mass matrix.

One of the factors that determines the flavor structure of a model is the relative orientation between the neutrino flavor eigenstates and the flavon VEVs. Assigning the three neutrinos as a triplet under the flavor group implies that they are aligned along the basis states of the representation. To obtain the alignment of flavon VEVs, we construct a flavon potential invariant under the discrete flavor group.  The fact that the symmetry is discrete limits the extremum points of the potential to a finite set. SSB randomly chooses one among these extrema as the vacuum alignment. By changing the nature of the flavon potential we may alter the set of extremum points and thus change the possible vacuum alignments. The flavon VEVs form the building blocks of the mass matrix, so the alignment of the VEVs in flavor space has important consequences for the structure of the mass matrix. We expect that a given alignment has specific symmetry properties under the flavor group, which in turn imparts specific features to the mass matrix.

\begin{figure}[]
\begin{center}
\includegraphics[trim={5cm 4cm 5cm 3cm},scale=0.45]{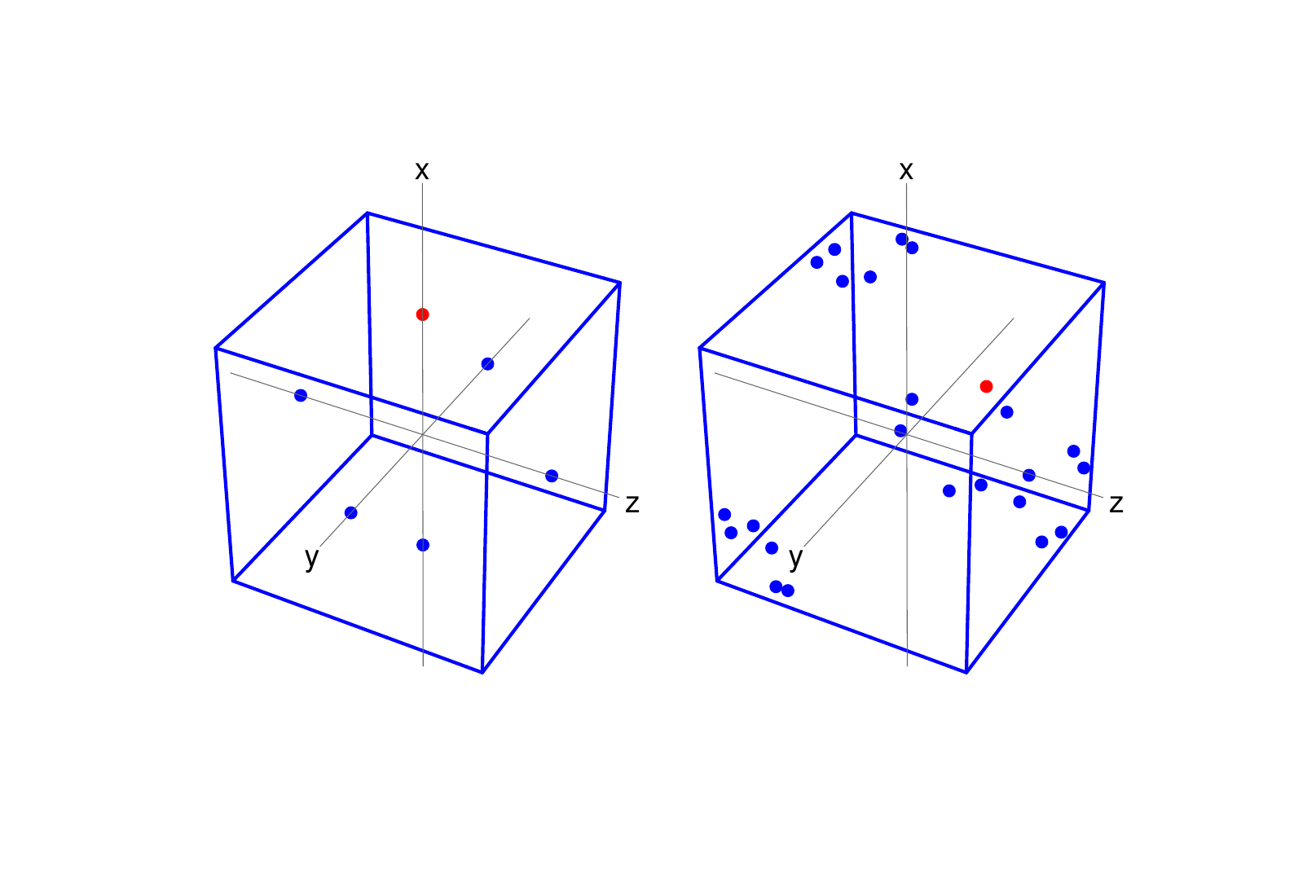}
\caption[The cubes]{The dots on the cubes represent extrema of two cases of flavon potentials which have $S_4$ symmetry (under $\rtp$). In the left figure, the extrema are on the face centers of the cube. The VEV, denoted by the red dot, is aligned along one of the axes of symmetries of the cube. In the right figure, $S_4$ symmetry of the potential results in $24$ extremum points positioned around the cube. However, the VEV (the red dot) is not aligned along any axis of symmetry.}
\label{fig:cubes}
\end{center}
\end{figure}

Let us use the discrete group $S_4$ as an example to study the alignment of states in the flavor space. The triplet representation ($\boldsymbol{3}$) of $S_4$ corresponds to the 24 proper rotations in three-dimensional real space that leaves a cube invariant. A convenient basis (e.g.,~\cite{Grimus:2010ak}) of $\boldsymbol{3}$ is where the basis states are aligned along the face centers of the cube, i.e., ~the x, y, and z axes as shown in Fig.~\ref{fig:cubes}. In this basis, we may use
\begin{equation}\label{eq:s4gen}
P \equiv
\left(\begin{matrix}0 & 0 & 1\\
       0 & -1 & 0\\
       1 & 0 & 0
\end{matrix}\right), \quad Q \equiv
\left(\begin{matrix}0 & 1 & 0\\
       0 & 0 & 1\\
       1 & 0 & 0
\end{matrix}\right)
\end{equation}
as the generators of the group. The group elements of $\boldsymbol{3}$ consist of 9 rotations (by angles $\pm \pi/2$, $\pi$) about the three axes passing through face centers, 8 rotations (by angles $\pm2\pi/3$)  about the four axes passing through vertices, 6 rotations (by an angle $\pi$) about the six axes passing through edge centers and the identity element. 

When we assume that the neutrinos transform as a $\boldsymbol{3}$,
\begin{equation}
\boldsymbol{3}\equiv\nu_R=(\nu_{R1},\nu_{R2},\nu_{R3})^T,
\end{equation}
we are assigning its components as the basis states of the triplet representation. In other words, $\nu_{R1}$, $\nu_{R2}$, and $\nu_{R3}$ correspond to $(1,0,0)^T$, $(0,1,0)^T$, and $(0,0,1)^T$, respectively. These states, since they are oriented along the face centers of the cube, have specific symmetry properties. They form the axes of symmetries of the cube about which rotations by multiples of $\pi/2$ keep the cube invariant. For example, the group element
\begin{equation}
PQ\equiv
\left(\begin{matrix}1 & 0 & 0\\
       0 & 0 & -1\\
       0 & 1 & 0
\end{matrix}\right)
\end{equation}
generates three rotations (by angles $\pm\pi/2$, $\pi$) about  the axis $(1,0,0)^T\equiv \nu_{R1}$. These rotations along with the identity element form a $C_4$ subgroup of $S_4$. Similarly we have two more $C_4$ subgroups in relation to the states $\nu_{R2}$ and $\nu_{R3}$. This shows that the neutrino states are uniquely defined by the subgroup structure of $S_4$.

The tensor product expansion of two triplets ($\boldsymbol{3}$) is given by
\begin{equation}\label{eq:expansion}
\boldsymbol{3}\otimes\boldsymbol{3}=\boldsymbol{1}\oplus\boldsymbol{2}\oplus\rtp\oplus\boldsymbol{3}.\\
\end{equation}
We couple two triplets of neutrinos using this expansion to obtain
\begin{align}
\boldsymbol{1}&\equiv(\nu_R\cdot\nu_R)_s=\nu_{R1}\cdot\nu_{R1}+\nu_{R2}\cdot\nu_{R2}+\nu_{R3}\cdot\nu_{R3},\label{eq:numul1}\\
\begin{split}
\hspace{-2mm}\boldsymbol{2}&\equiv(\nu_R\cdot\nu_R)_d=(2\nu_{R1}\cdot\nu_{R1}-\nu_{R2}\cdot\nu_{R2}-\nu_{R3}\cdot\nu_{R3},\\
&\quad\quad\quad\quad\quad\quad\quad\quad\quad\sqrt{3}\nu_{R2}\cdot\nu_{R2}-\sqrt{3}\nu_{R3}\cdot\nu_{R3})^T,
\end{split}\label{eq:numul2}\\
\begin{split}
\hspace{-2mm}\rtp&\equiv(\nu_R\cdot\nu_R)_t=(\nu_{R2}\cdot\nu_{R3}+\nu_{R3}\cdot\nu_{R2},\\ 
&\quad\quad\quad\nu_{R3}\cdot\nu_{R1}+\nu_{R1}\cdot\nu_{R3}, \,\,\nu_{R1}\cdot\nu_{R2}+\nu_{R2}\cdot\nu_{R1})^T.
\end{split}\label{eq:numul3}
\end{align}
The $\boldsymbol{3}$ in the right-hand side (RHS) of Eq.~(\ref{eq:expansion}) is antisymmetric and hence it vanishes. The generators, corresponding to Eqs.~(\ref{eq:s4gen}), for the doublet ($\boldsymbol{2}$) and the triplet ($\rtp$) are
\begin{equation}\label{eq:s4dblt}
P \equiv
\left(\begin{matrix}-\frac{1}{2} & -\frac{\sqrt{3}}{2}\\
       -\frac{\sqrt{3}}{2} & \frac{1}{2}
\end{matrix}\right), \quad Q \equiv
\left(\begin{matrix}-\frac{1}{2} & \frac{\sqrt{3}}{2}\\
       -\frac{\sqrt{3}}{2} & -\frac{1}{2}
\end{matrix}\right)
\end{equation}
and
\begin{equation}\label{eq:3prime}
P \equiv
\left(\begin{matrix}0 & 0 & -1\\
       0 & 1 & 0\\
       -1 & 0 & 0
\end{matrix}\right), \quad Q \equiv
\left(\begin{matrix}0 & 1 & 0\\
       0 & 0 & 1\\
       1 & 0 & 0
\end{matrix}\right),
\end{equation}
respectively, where the basis adopted is as per the C-G coefficients given in Eqs.~(\ref{eq:numul2}) and (\ref{eq:numul3}). 

For the purpose of this discussion, we introduce a flavon triplet $\phi=(\phi_1, \phi_2, \phi_3)^T$ which transforms as a $\rtp$. It is coupled to $(\nu_R\cdot\nu_R)_t$, Eq.~(\ref{eq:numul3}), to obtain the invariant mass term,
\begin{equation}
(\nu_R\cdot\nu_R)_t^T \phi.
\end{equation}
The flavon acquires the VEV, $\langle\phi\rangle=(\langle\phi_1\rangle,\langle\phi_2\rangle,\langle\phi_3\rangle)^T$, through SSB. The mass matrix is obtained in terms of this VEV,
\begin{equation}
(\nu_R\cdot\nu_R)_t^T \langle\phi\rangle = \left(\begin{matrix}\nu_{R1}\\
	\nu_{R2}\\
\nu_{R3}
\end{matrix}\right)^T \left(\begin{matrix}0 & \langle\phi_3\rangle & \langle\phi_2\rangle\\
       \langle\phi_3\rangle & 0 & \langle\phi_1\rangle\\
       \langle\phi_2\rangle & \langle\phi_1\rangle& 0
\end{matrix}\right)\cdot\left(\begin{matrix}\nu_{R1}\\
	\nu_{R2}\\
\nu_{R3}
\end{matrix}\right).
\end{equation}

The representation $\rtp$ also corresponds to 24 rotational symmetries of the cube; however, 12 of them are improper rotations. The proper rotations consist of 3 rotations (by an angle $\pi$) about the three axes passing through face centers, 8 rotations (by angles $\pm2\pi/3$) about the four axes passing through vertices and the identity element. The improper rotations consist of 6 rotations (by angles $\pm\pi/2$) about the three axes passing through face centers and 6 rotations (by an angle $\pi$) about the six axes passing through edge centers combined with space inversion.

Suppose we construct a flavon potential that has an extremum point at $\phi=(1,0,0)^T$. Because of $S_4$ symmetry, the potential will have similar extrema at all points generated by the action of $S_4$ on $\phi=(1,0,0)^T$. There are six such points,
\begin{equation}\label{eq:res1}
(\pm1,0,0), (0,\pm1,0), (0,0,\pm1),
\end{equation} 
corresponding to the six face centers of the cube as shown in Fig.~\ref{fig:cubes} (left). Note that, even though the representation $\rtp$ has 24 distinct elements, the action of those elements produces only six distinct points. This is explained using the orbit-stabilizer theorem. The orbit of a point is defined as the set of all points obtained by the group action on that given point; i.e.,~Eq.(\ref{eq:res1}) forms the orbit of $(1,0,0)^T$. The stabilizer of a point is defined as the set of all group elements under whose action the given point remains invariant. $(1,0,0)^T$ remains invariant under the action of 
\begin{equation}\label{eq:genc2c2}
(PQ)^2 \equiv
\left(\begin{matrix}1 & 0 & 0\\
       0 & -1 & 0\\
       0 & 0 & -1
\end{matrix}\right), \quad QPQ^2 \equiv
\left(\begin{matrix}1 & 0 & 0\\
       0 & 0& -1\\
       0 & -1 & 0
\end{matrix}\right).
\end{equation}
$(PQ)^2$ and $QPQ^2$ generate $C_2$ groups individually, and they commute with each other. So they generate a $C_2\times C_2$ subgroup of $S_4$. In other words, the stabilizer of $(1,0,0)^T$  is the $C_2\times C_2$ subgroup generated by $(PQ)^2$ and $QPQ^2$. The orbit-stabilizer theorem states that 
\begin{equation}
|\text{Orb}(x)| = \frac{|G|}{|\text{Stab}(x)|},
\end{equation}
where $|\text{Orb}(x)|$ is the number of points in the orbit of $x$, $|G|$ is the number of elements in the group and $|\text{Stab}(x)|$ is the number of elements in the stabilizer of $x$. In our case, the stabilizer ($C_2\times C_2$) has four elements. Therefore we obtain the number of points in the orbit (the number of extrema of the potential) to be $24/4=6$, consistent with Eq.~(\ref{eq:res1}). When the flavor symmetry group ($S_4$) is broken through the mechanism of SSB, the resulting vacuum alignment will be along one of these extrema. So the symmetry breaking will not be complete; the $C_2 \times C_2$ stabilizer subgroup of the given VEV will remain as the unbroken residual symmetry. 

We may also construct a potential that has an extremum at the point $(1,1,1)^T$. The orbit for this point is
\begin{equation}\label{eq:res2a}
(1,1,1), (1,-1,-1), (-1,1,-1), (-1,-1,1).
\end{equation}
These points form one set of four nonopposing vertices of the cube and they constitute a tetrahedron. The stabilizer of $(1,1,1)^T$ is generated by
\begin{equation}\label{eq:vertexstab}
Q \equiv
\left(\begin{matrix}0 & 1 & 0\\
       0 & 0 & 1\\
       1 & 0 & 0
\end{matrix}\right), \quad PQ^2PQP \equiv
\left(\begin{matrix}1 & 0 & 0\\
       0 & 0& 1\\
       0 & 1 & 0
\end{matrix}\right).
\end{equation}
$Q$ and $PQ^2PQP$ generate $C_3$ and $C_2$, respectively, and taken together they generate the dihedral group $D_6$ which forms the stabilizer of $(1,1,1)^T$. The number of elements in the orbit is $24/6=4$, consistent with Eq.~(\ref{eq:res2a}). We also have a second set of four nonopposing vertices,
\begin{equation}\label{eq:res2b}
(-1,-1,-1), (-1,1,1), (1,-1,1), (1,1,-1),
\end{equation}
which form another orbit. A VEV aligned along a point in one of these orbits, Eqs.~(\ref{eq:res2a}) and (\ref{eq:res2b}), breaks $S_4$ into the corresponding $D_6$ subgroup which would remain as the unbroken residual symmetry. The orbits we have discussed, Eqs.~({\ref{eq:res1}), (\ref{eq:res2a}), and (\ref{eq:res2b}), are the only ones that are uniquely defined in terms of the symmetries of the $\rtp$ of $S_4$.

It is also possible to construct a potential whose extrema are oriented in directions not defined by any symmetry. In general, such a potential will have $24$ distinct extrema as shown in Fig.~\ref{fig:cubes} (right). These points also form an orbit, but it is not uniquely defined unlike the orbits discussed earlier. In other words, the stabilizers of points in such an orbit are trivial subgroups.\footnote{The stabilizer of a point being nontrivial does not always ensure that the orbit of the point is unique. Consider the point $(\alpha,1,1)^T$. This point remains invariant under the action of the $C_2$ subgroup generated by $PQ^2PQP$, Eq.~(\ref{eq:vertexstab}). This $C_2$ subgroup forms the nontrivial stabilizer of $(\alpha,1,1)^T$. It is clear that this orbit is not unique; rather it depends on the arbitrary parameter $\alpha$.} A VEV along one of these extrema breaks $S_4$ completely so that there remains no residual symmetry. By appropriately tuning the potential, we will be able to orient the extremum points and the resulting VEV in almost any direction we may want. This is true for all discrete groups, not just $S_4$. For explicit construction of flavon potentials of $S_4$, refer to Appendix~A.

When the VEV as shown in Fig.~\ref{fig:cubes} (right) breaks $S_4$ symmetry, we may be tempted to conclude that there is a 1 in 24 chance of obtaining such a choice. Even though this is true for a given potential, the directions of its 24 minima can be continuously tuned using the parameters appearing in the potential. As a result, the net probability of obtaining the VEV along a specific direction is vanishingly small (1 in infinity). This is unlike the case of the VEV shown in Fig.~\ref{fig:cubes} (left). Here, varying the parameters in the potential will not continuously alter the directions of the six minima of the potential, because these directions are fully defined by the structure of the discrete group itself. The net probability of obtaining the given VEV, Fig.~\ref{fig:cubes} (left), remains to be 1 in 6.

We note that a considerable number of publications rely almost entirely on flavon potentials to determine their vacuum alignments. Authors utilize quite complicated potentials to obtain VEVs that are phenomenologically viable, but they fail to provide a justification for these VEVs in terms of the symmetries of the flavor group. Even though this procedure is technically valid, we argue that it goes against the very spirit of using the properties of the discrete groups for determining the flavor structure. If the VEV is made to orient in an arbitrary direction with no apparent connection to the original symmetry, the whole purpose of using discrete symmetries can be called into question. We argue that the orientations of the neutrino (fermion) basis states as well as the flavon VEVs should be uniquely specifiable in terms of the residual symmetries that form subgroups of the discrete flavor group. The mathematical elegance of the subgroup structure of the flavor group should manifest as the restrictiveness of the orientations of the flavor states and thus the predictiveness of the flavor model.

Flavor models generally involve several irreducible multiplets of flavons. We can assign different discrete charges to these multiplets so that they are decoupled in the flavon potential.\footnote{A detailed discussion of how this decoupling can be achieved is beyond the scope of this work.} In such a scenario, the vacuum alignment of one multiplet can be chosen independently of the others. For example, an $S_4$ invariant potential can be constructed with two decoupled irreducible triplets ($\rtp$). One of them may obtain a VEV $\propto(\pm1,0,0)$ which is fully defined by a $C_2\times C_2$ residual symmetry of $S_4$ and which has a probability of 1 in 3. The other triplet may obtain a VEV $\propto(0, \pm1, 0)$ which is fully defined by another $C_2\times C_2$ symmetry and which also has a probability of 1 in 3. We have a total probability of 1 in 9 to obtain both of these VEVs. Taken together, the two VEVs fully break $S_4$ and we are left with no residual symmetries. However, we argue that this situation is quite different from the case where a single irreducible multiplet fully breaks $S_4$ as shown in Fig.~\ref{fig:cubes} (right). In that case, the probability is vanishingly small. 

When the flavon vacuum alignments, Eqs.~(\ref{eq:vevtpmp}) and (\ref{eq:vevtpmm}), for the sextet of $\sgm$ were proposed~\cite{Krishnan:2018tja}, they were not uniquely defined using their symmetry properties. In this paper, we combine $\sgm$ with a new discrete symmetry group which we call $\bt$. We introduce flavons which transform under both $\sgm$ and $\bt$. Their VEVs uniquely break the combined flavor group into its subgroups; i.e.,~the VEVs are completely determined by their residual symmetries alone. These flavons are coupled together to obtain the sextet of $\sgm$. This sextet in turn couples with the neutrino triplets resulting in the Majorana mass term. In this work, we do not construct flavon potentials. Instead, we follow the arguments presented in this section to obtain the VEVs; i.e., all the flavon irreducible multiplets are assigned VEVs which are fully defined in terms of their respective residual symmetries.

\section{The Discrete Group $\bt$}

We construct discrete group, $\bt$, using the following generators:
\begin{equation}\label{eq:gen}
\A =
\left(\begin{matrix}
	0 & 0 & 0 & 0 & 1 & 0\\
     	 0 & 0 & 0 & 0 & 0 & 1\\
      	0 & 0 & 0 & 1 & 0 & 0\\
	0 & 1 & 0 & 0 & 0 & 0\\
	0 & 0 & 1 & 0 & 0 & 0\\
	1 & 0 & 0 & 0 & 0 & 0
\end{matrix}\right), \quad  \B =
\left(\begin{matrix}
	\om & 0 & 0 & 0 & 0 & 0\\
     	0 & 0 & \Ta & 0 & 0 & 0\\
      	0 & \Tb & 0 & 0 & 0 & 0\\
	0 & 0 & 0 & 0 & \Tb & 0\\
	0 & 0 & 0 & \Ta & 0 & 0\\
	0 & 0 & 0 & 0 & 0 & \ob
\end{matrix}\right),
\end{equation}
where $\om=e^{i\frac{2\pi}{3}}$ and $\ob=e^{-i\frac{2\pi}{3}}$ are the cube roots of unity and $\Ta=e^{i\frac{\pi}{4}}$ and $\Tb=e^{-i\frac{\pi}{4}}$ are the eighth roots of unity. The largest cyclic subgroup of this group is $C_{24}$, generated by $\om \Ta$ and hence the subscript $24$ in $\bt$. These generators, Eq.~(\ref{eq:gen}),  are selected so that the group constructed from them helps to uniquely define the required flavon VEVs. The rest of this section covers the mathematical study of the properties of this group. A reader who is more inclined toward applying the group theoretical results for the construction of the VEVs and the mass matrix may skip over to Sec.~4 and may revert to this section when it is deemed necessary. 

As the first step in analyzing $\bt$, we construct the group elements,
\begin{align}
\ceo &= (\A\!^2\,\B\,\A\!^3\,\B\,\A\!^3)^9=\text{Diag}(1,\Tb,\Ta,1,1,1),\label{eq:ceo}\\
\cet &= (\A\!^2\,\B\,\A\!^5\,\B\,\A)^9=\text{Diag}(1,1,1,\Ta,\Tb,1). \label{eq:cet}
\end{align}
Using $\ceo$, $\cet$ and $\B$ we obtain the group element,
\begin{equation}\label{eq:genbm}
\Bm = \ceo\,\cet\,\B^3 =
\left(\begin{matrix}
	1 & 0 & 0 & 0 & 0 & 0\\
     	0 & 0 & 1 & 0 & 0 & 0\\
      	0 & 1 & 0 & 0 & 0 & 0\\
	0 & 0 & 0 & 0 & 1 & 0\\
	0 & 0 & 0 & 1 & 0 & 0\\
	0 & 0 & 0 & 0 & 0 & 1
\end{matrix}\right).
\end{equation}
$\A$ and $\Bm$ generate the group $S_3\times S_3$ which forms a subgroup of $\bt$. To show this we obtain
\begin{align}
\Do &=
\left(\begin{matrix}
	0 & 0 & 0 & 0 & 0 & 1\\
     	0 & 0 & 0 & 0 & 1 & 0\\
      	0 & 0 & 0 & 1 & 0 & 0\\
	0 & 0 & 1 & 0 & 0 & 0\\
	0 & 1 & 0 & 0 & 0 & 0\\
	1 & 0 & 0 & 0 & 0 & 0
\end{matrix}\right), &&\Eo =
\left(\begin{matrix}
	0 & 0 & 1 & 0 & 0 & 0\\
     	1 & 0 & 0 & 0 & 0 & 0\\
      	0 & 1 & 0 & 0 & 0 & 0\\
	0 & 0 & 0 & 0 & 0 & 1\\
	0 & 0 & 0 & 1 & 0 & 0\\
	0 & 0 & 0 & 0 & 1 & 0
\end{matrix}\right),\label{eq:s31gen}\\
\Dt & =
\left(\begin{matrix}
	0 & 0 & 0 & 1 & 0 & 0\\
     	0 & 0 & 0 & 0 & 1 & 0\\
      	0 & 0 & 0 & 0 & 0 & 1\\
	1 & 0 & 0 & 0 & 0 & 0\\
	0 & 1 & 0 & 0 & 0 & 0\\
	0 & 0 & 1 & 0 & 0 & 0
\end{matrix}\right), &&\Et =
\left(\begin{matrix}
	0 & 0 & 1 & 0 & 0 & 0\\
     	1 & 0 & 0 & 0 & 0 & 0\\
      	0 & 1 & 0 & 0 & 0 & 0\\
	0 & 0 & 0 & 0 & 1 & 0\\
	0 & 0 & 0 & 0 & 0 & 1\\
	0 & 0 & 0 & 1 & 0 & 0
\end{matrix}\right),\label{eq:s32gen}
\end{align}
where
\begin{align}
&\Do = \A\!^2\,(\A\,\Bm)^3, &&\Eo = \A\!^2,\label{eq:s31genexp}\\
&\Dt = \A\!^3, &&\Et = (\A\,\Bm)^2.\label{eq:s32genexp}
\end{align}
$\Do, \Eo$ and $\Dt, \Et$ in Eqs.~(\ref{eq:s31gen}) and (\ref{eq:s32gen}) separately form the generators of the group $S_3$, because they satisfy the following group presentation:
\begin{equation}
\langle \Di,\Ei ~|~ \Di\!^2 = \Ei\!^3 = (\Di\,\Ei)^2 = {\bf1} \rangle ~\text{for}~ i=1,2,
\end{equation}
along with the relationship,
\begin{equation}
\Ei\,\Di = \Di\,\Ei\!^2 ~\text{for}~ i=1,2.
\end{equation}
$S_3$ group elements, $g_1$ and $g_2$, generated by $\Do, \Eo$ and $\Dt, \Et$, respectively, can be expressed as 
\begin{equation}
g_1=\Do\!^{i_1}\,\Eo\!^{j_1}, \quad g_2=\Dt\!^{i_2}\,\Et\!^{j_2},
\end{equation} 
where $i_1, i_2 \in \{1, 2\}$ and $j_1, j_2 \in \{1, 2,3\}$. The first set of generators, Eq.~(\ref{eq:s31gen}), commutes with the second set, Eq.~(\ref{eq:s32gen}); i.e.,
\begin{equation}
[ \Di,\Dj] = [ \Ei,\Ej] = [ \Di,\Ej]  = 0  ~\text{for}~ i \neq j,
\end{equation}
so that we obtain the direct product of two $S_3$ groups. Thus we show that $\A$ and $\Bm$ generate the group $S_3\times S_3$ with the total number of elements equal to $2\times3\times2\times3$. Note that the elements of $S_3\times S_3$ in the basis given by Eqs.~(\ref{eq:s31gen}) and (\ref{eq:s32gen}), are matrices with 1's and 0's only.

$\ceo$ and $\cet$, Eqs.~(\ref{eq:ceo}) and (\ref{eq:cet}), individually generate the cyclic group $C_8$. In $\bt$, we can find two more such generators of $C_8$,
\begin{align}
\cer &= \Eo\,\ceo\,\Eo\!^2=\text{Diag}(\Ta,1,\Tb,1,1,1),\label{eq:cer}\\
\cef &= \Et\,\cet\,\Et\!^2=\text{Diag}(1,1,1,\Tb,1,\Ta). \label{eq:cef}
\end{align}
Four elements, similar to Eqs.~(\ref{eq:ceo}), (\ref{eq:cet}), (\ref{eq:cer}), and (\ref{eq:cef}), which individually generate the cyclic group $C_3$ can also be found:
\begin{align}
\cro &= \Dt\B^2\Dt\Eo\B^2\Eo^2=\text{Diag}(1,\ob,\om,1,1,1),\label{eq:cro}\\
\crt &= \Do\cro\Do=\text{Diag}(1,1,1,\om,\ob,1), \label{eq:crt}\\
\crr &= \Eo\cro\Eo^2=\text{Diag}(\om,1,\ob,1,1,1),\label{eq:crr}\\
\crf &= \Et\crt\Et^2=\text{Diag}(1,1,1,\ob,1,\om). \label{eq:crf}
\end{align}
We also find a fifth independent $C_3$ generator,
\begin{equation}
\crv = \B^2=\text{Diag}(\ob,1,1,1,1,\om).\label{eq:crv}
\end{equation}
Using $6 \times 6$ special unitary diagonal matrices, the maximum number of independent $C_n$ generators that can be constructed is five, and in Eqs.~(\ref{eq:cro})--(\ref{eq:crv}) we have listed all of them for $C_3$. For the case of the diagonal $C_8$ subgroups of $\bt$, it so happens that the upper and the lower $3\times3$ diagonal matrices are individually special unitary. This additional constraint limits the total number of independent generators to four, i.e., Eqs.~(\ref{eq:ceo}), (\ref{eq:cet}), (\ref{eq:cer}), and (\ref{eq:cef}). Equations~(\ref{eq:ceo}), (\ref{eq:cet}), (\ref{eq:cer}), (\ref{eq:cef}), and (\ref{eq:cro})--(\ref{eq:crv}) constitute an exhaustive list of generators producing all the diagonal elements within $\bt$. These elements form the subgroup $C_8\times C_8\times C_8\times C_8\times C_3\times C_3\times C_3\times C_3\times C_3$ of $\bt$. The diagonal elements commute with each other, and they form the largest Abelian subgroup of $\bt$. Note that $3$ and $8$ are co-prime numbers that implies $C_8\times C_3$ is $C_{24}$. This can also be inferred from the multiplication of $C_8$ and $C_3$ generators, for example,
\begin{equation}
 \ceo \cro=\text{Diag}(1,\Tb\ob,\Ta\om,1,1,1)=\ctw.\label{eq:cto}
\end{equation}
In other words, the group $C_{24}\times C_{24}\times C_{24}\times C_{24}\times C_3$ forms the largest Abelian subgroup of $\bt$. 

Every representation matrix of $\bt$ is of the form of a representation matrix of $S_3\times S_3$ with phases replacing certain number of $1$'s in the $S_3\times S_3$ matrix. These phases can be extracted out using a diagonal phase matrix, i.e.~an element of the aforementioned Abelian subgroup. In other words, any element of $\bt$ can be obtained by left multiplying (or right multiplying) the corresponding element of $S_3\times S_3$ with an appropriate diagonal phase matrix. Therefore, $C_{24}\times C_{24}\times C_{24}\times C_{24}\times C_3$ and $S_3\times S_3$ form a normal subgroup and the associated quotient group, respectively, of $\bt$. Using this information, we may express $\bt$ as a semidirect product,
\begin{equation}\label{eq:semidirect}
\bt = (C_{24}\times C_{24}\times C_{24}\times C_{24}\times C_3) \rtimes (S_3\times S_3).
\end{equation}
Any element of $\bt$ can be uniquely expressed as
\begin{equation}\label{eq:unique}
\begin{split}
g&= \ceo^{\!\!\!m_1}\cet^{\!\!\!m_2}\cer^{\!\!\!m_3}\cef^{\!\!\!m_4}\\
&\quad\quad\times\cro^{\!\!\!n_1}\crt^{\!\!\!n_2}\crr^{\!\!\!n_3}\crf^{\!\!\!n_4}\crv^{\!\!\!n_5}\\
&\quad\quad\quad\times\Do\!^{i_1}\,\Eo\!^{j_1}\,\Dt\!^{i_2}\,\Et\!^{j_2},
\end{split}
\end{equation} 
where $m_1, ..., m_4 \in \{1, ..., 8\}$;  $n_1, ..., n_5 \in \{1, 2,3\}$; $i_1, i_2 \in \{1,2\}$; and $j_1, j_2 \in \{1,2,3\}$. So, it is clear that the order of the group $\bt$ is $8^4\,3^5\,2^2\,3^2$. We used the group theory package GAP~\cite{GAP4} and verified that the group generated by $\A$ and $\B$ has this order, thus confirming our calculations. We also verified that the sextet representation, Eqs.~(\ref{eq:gen}), is irreducible. We note that $\bt$ is not a subgroup of $U(3)$.

\section{The Model}

The complete flavor group for our model is $G_f=\sgm\times C_3 \times C_4 \times \bt\times\bt$. This is of the form $G_f=G_r\times G_x$ where $G_r=\sgm \times C_3\times C_4$ and $G_x=\bt\times\bt$. $G_x$, which we call the auxiliary group, is defined as the group under which only the flavons transform nontrivially. This group provides additional symmetries to the flavons and helps us to define their VEVs in terms of these symmetries. Fermions are invariant singlets under $G_x$. On the other hand, $G_r$ is defined as the group under which both fermions and flavons are allowed to transform nontrivially. Table~\ref{tab:flavourcontent} shows how the fermion and the flavon fields transform under $G_f$. The $C_3$ group helps in the construction of the charged-lepton mass term using the flavons $\phi_\mu$ and $\phi_\tau$. The neutrino Majorana mass term is constructed using the flavons $\phia$, $\phib$ and $\bind$. The $C_4$ group is added so that $\phi_\mu$ and $\phi_\tau$ do not couple in the Majorana sector.

The mass term in the charged-lepton sector is given by
\begin{equation}\label{eq:cllag}
y_\mu \bar{L} \frac{\phi_\mu}{\Lambda} \mu_R H+y_\tau \bar{L} \frac{\phi_\tau}{\Lambda} \tau_R H + y_e \bar{L} \frac{(\bar{\phi}_\mu\bar{\phi}_\tau)_{\boldsymbol{\tb}}}{\Lambda^2} e_R H
\end{equation}
where
\begin{equation}\label{eq:combmutau}
\begin{split}
&(\bar{\phi}_\mu\bar{\phi}_\tau)_{\boldsymbol{\tb}}\\&= (\bar{\phi}_{\mu 2}\bar{\phi}_{\tau 3}-\bar{\phi}_{\mu 3}\bar{\phi}_{\tau 2}, \bar{\phi}_{\mu 3}\bar{\phi}_{\tau 1}-\bar{\phi}_{\mu 1}\bar{\phi}_{\tau 3}, \bar{\phi}_{\mu 1}\bar{\phi}_{\tau 2}-\bar{\phi}_{\mu 2}\bar{\phi}_{\tau 1})^T
\end{split}
\end{equation}
transforms as a $\boldsymbol{\tb}$ under $\sgm$.

The flavons $\phi_\tau$ and $\phi_\mu$ are assigned the following vacuum alignments:
\begin{equation}\label{eq:vevmutau}
\langle \phi_\mu \rangle = v_\mu (1, \ob, \om)^T, \quad \langle \phi_\tau \rangle = v_\tau (1, \om, \ob)^T.
\end{equation}
Equations~(\ref{eq:combmutau}) and (\ref{eq:vevmutau}) lead to 
\begin{equation}
\langle (\bar{\phi}_\mu\bar{\phi}_\tau)_{\boldsymbol{\tb}}\rangle =  v_\mu v_\tau i\sqrt{3} (1, 1, 1)^T.
\end{equation}
Substituting the flavon VEVs along with the Higgs VEV, $ \langle H \rangle=(0, v)^T$, in the Lagrangian, Eq.~(\ref{eq:cllag}), we obtain the charged-lepton mass term after spontaneous symmetry breaking,
\begin{equation}
\bar{l}_LM_ll_R 
\end{equation}
where 
\begin{equation}
l_L=(e_L, \mu_L, \tau_L)^T, \quad l_R=(e_R, \mu_R, \tau_R)^T
\end{equation}
and 
\begin{equation}
M_l =i\sqrt{3} \frac{vy_ev_\mu v_\tau}{\Lambda^2}\left(\begin{matrix} 1 & 0 & 0\\
1 & 0 & 0\\
1 & 0 & 0
\end{matrix}\right)+ \frac{v}{\Lambda}\left(\begin{matrix} 0 & y_\mu v_\mu & y_\tau v_\tau\\
0 & \ob y_\mu v_\mu & \om y_\tau v_\tau\\
0 & \om y_\mu v_\mu & \ob y_\tau v_\tau
\end{matrix}\right).
\end{equation}
$M_l$ is diagonalized using the unitary matrix $V$, Eq.~(\ref{eq:gen3}), as follows:
\begin{equation}
V M_l \text{diag}(1, i, i) = \text{diag}(m_e, m_\mu, m_\tau)
\end{equation}
where $m_e = 3y_e v \frac{v_\mu v_\tau}{\Lambda^2}$, $m_\mu=\sqrt{3}y_\mu v \frac{v_\mu}{\Lambda}$ and $m_\tau=\sqrt{3}y_\tau v \frac{v_\tau}{\Lambda}$ are the charged-lepton masses. The diagonalizing matrix, $V$, is the same as that obtained in Ref.~\cite{Krishnan:2018tja}.

{\renewcommand{\arraystretch}{1.5}
\begin{table}[]
\begin{center}
\begin{tabular}{|c| c c c c c c c c c c|}
\hline
&$e_R$&$\mu_R$&$\tau_R$&$L$&$\phi_\mu$&$\phi_\tau$&$\nu_R$&$\phia$&$\phib$&$\bind$\\
\hline
$\sgm$&$\boldsymbol{1}$&$\boldsymbol{1}$&$\boldsymbol{1}$&$\boldsymbol{\tb}$&$\boldsymbol{\tb}$&$\boldsymbol{\tb}$&$\boldsymbol{\tb}$&$\boldsymbol{3}$&$\boldsymbol{3}$&$\boldsymbol{1}$\\
$C_3 \times C_4$&$\boldsymbol{i}$&$\boldsymbol{\om i}$&$\boldsymbol{\ob i}$&$\boldsymbol{i}$&$\boldsymbol{\ob}$&$\boldsymbol{\om}$&$\boldsymbol{i}$&$\boldsymbol{1}$&$\boldsymbol{1}$&$\boldsymbol{-1}$\\
$\bt\times\bt$&$\boldsymbol{1}$&$\boldsymbol{1}$&$\boldsymbol{1}$&$\boldsymbol{1}$&$\boldsymbol{1}$&$\boldsymbol{1}$	&$\boldsymbol{1}$&$\boldsymbol{6}\times\boldsymbol{1}$&$\boldsymbol{1}\times\boldsymbol{6}$&$\boldsymbol{\bar{6}}\times\boldsymbol{\bar{6}}$\\
\hline
\end{tabular}
\caption{The flavor structure of the model.}
\label{tab:flavourcontent}
\end{center}
\end{table}
}

For the neutrinos, the Dirac mass term is given by
\begin{equation}
y_\nu \bar{L} \nu_R \tilde{H}.
\end{equation}
This term leads to a diagonal mass matrix as given in Ref.~\cite{Krishnan:2018tja}. The rest of this section deals with the construction of the Majorana mass term using the flavons $\phia$, $\phib$, and $\Delta$. The flavons $\phia$ and $\phib$ are triplets under $\sgm$ and sextets under the first and the second copies of $\bt$ respectively, Table~\ref{tab:flavourcontent}. The flavon $\bind$ transforms as a conjugate sextet under both copies of $\bt$. In this paper, we use Latin and Greek letters to denote the indices that transform under $\sgm$ and $\bt$, respectively, e.g.,~$\nu_{Ri}$, $\phia_{\alpha i}$, $\phib_{\alpha i}$, $\bind_{\alpha\beta}$. 

Using the C-G coefficients, Eqs.~(\ref{eq:cg}), and considering the transformation properties given in Table~\ref{tab:flavourcontent}, we construct the invariant term in the Majorana sector,
\begin{equation}\label{eq:majmass}
\mathcal T_\text{Maj} = \sum S_{ijk}S_{imn}\,\,\phia_{\alpha m}\phib_{\beta n} \bind_{\alpha\beta} \,\,\nu_{Rj}\cdot\nu_{Rk},
\end{equation}
where the summation is over all repeated indices. Comparing this invariant with Eq.~(\ref{eq:Tnu}), we obtain
\begin{equation}\label{eq:coupled}
\xi_i=\sum S_{imn}\phia_{\alpha m}\phib_{\beta n} \bind_{\alpha\beta}.
\end{equation}
The flavon $\phia$ (and $\phib$) can be considered as a set of six $\sgm$ triplets. In Eq.~(\ref{eq:coupled}), we have a composite system of these triplets coupled together with $\bind$ to obtain $\xi$ which is a sextet under $\sgm$ and an invariant singlet under $\bt$. 

The flavons $\phia$, $\phib$, and $\bind$ acquire VEVs through SSB. Let these vacuum alignments be
\begin{equation}\label{eq:flavvevs}
\langle \phia \rangle = \langle \phib \rangle \propto
\left(\begin{matrix}
	\Ta & 0 & 0\\
     	0 & \om & 0\\
      	0 & 0 & \Tb\\
	-i & 0 & 0\\
	0 & \ob & 0\\
	0 & 0 & i
\end{matrix}\right), \langle \bind \rangle \propto
\left(\begin{matrix}
	1 & 0 & 0 & 1 & 0 & 0\\
     	0 & 1 & 0 & 0 & 1 & 0\\
      	0 & 0 & 1 & 0 & 0 & 1\\
	0 & 0 & 1 & 0 & 0 & 1\\
	0 & 1 & 0 & 0 & 1 & 0\\
	1 & 0 & 0 & 1 & 0 & 0
\end{matrix}\right).
\end{equation}
Here we have listed the components of the flavons with the help of matrices. The rows and the columns of $\bind$ denote the indices of the first and the second copies of $\bt$ in the flavor group. The rows and the columns of $\phia$ ($\phib$) denote the first (second) $\bt$ index and the $\sgm$ index respectively. Substituting the values of the VEVs, Eqs.~(\ref{eq:flavvevs}), in the expression for $\xi_i$, Eq.~(\ref{eq:coupled}), we obtain $\langle\xi\rangle$ as given in Eq.~(\ref{eq:vevtpmp}). On the other hand, if we use the conjugates of Eqs.~(\ref{eq:flavvevs}) we obtain $\langle\xi\rangle$ as given in Eq.~(\ref{eq:vevtpmm}). The mass term, Eq.~(\ref{eq:majmass}), can be written as a matrix equation,
\begin{equation}
\mathcal T_\text{Maj}=\nu_R^T \,\frac{1}{2}\left(\phia^T \bind\phib+\phib^T\bind^T\phia\right)\cdot\,\nu_R.
\end{equation}
Consequently, the expression for the $3\times3$ Majorana mass matrix becomes
\begin{equation}\label{eq:majexpr}
 M_\text{Maj}=\frac{1}{2}\left(\langle \phia\rangle^T \langle\bind\rangle\langle\phib\rangle+\langle\phib\rangle^T\langle\bind\rangle^T\langle\phia\rangle\right).
\end{equation}
Substituting the flavon VEVs, Eqs.~(\ref{eq:flavvevs}), (or their conjugates) in Eq.~(\ref{eq:majexpr}), we obtain the Majorana mass matrices, Eqs.~(\ref{eq:tpmmat}). 
Using this Majorana mass matrix and the diagonal mass matrix in the Dirac sector we can construct the type-1 seesaw mass matrix. The product of the diagonalizing matrices of the charged-lepton mass matrix and the seesaw mass matrix gives rise to $\tpm$ mixing with $\phi=\pm\pi/16$. This proceeds in the same way as given in Ref.~\cite{Krishnan:2018tja}. For a discussion of the resulting phenomenology, the reader may go through the aforementioned reference.

It may be noted that the VEVs ($\langle\phia\rangle$, $\langle\phib\rangle$, $\langle\bind\rangle$), Eqs.~(\ref{eq:flavvevs}), are composed of zeros and complex roots of unity. These VEVs are analogous to the points of residual symmetries on the cube, Eqs.~(\ref{eq:res1}), (\ref{eq:res2a}), and (\ref{eq:res2b}), which are composed of zeros and the two real roots of unity ($\pm1$). The points on the cube live in three-dimensional real space while the flavon VEVs live in higher dimensional (18, 18, and 36) complex spaces. In the next section, we prove that $\langle\phia\rangle$, $\langle\phib\rangle$, and $\langle\bind\rangle$ are indeed points having residual symmetries. Their orbits are uniquely defined with specific subgroups of $\sgm\times\bt\times\bt$ forming the stabilizer groups.

\section{The flavon vacuum alignments}

Consider the VEV, $\langle \phi_\mu \rangle$, Eq.~(\ref{eq:vevmutau}). It remains invariant under the group action,
\begin{equation}
\om E \langle \phi_\mu \rangle= \langle \phi_\mu \rangle
\end{equation}
The group element $\om E$ generates a $C_3$ group, $\{\om E, \ob E^2, 1\}$. The vacuum alignment, $\langle \phi_\mu \rangle$, breaks the flavor group into this $C_3$ residual group which, in turn, uniquely defines the alignment. Similarly, the VEV, $\langle \phi_\tau \rangle$, is uniquely defined by another $C_3$ subgroup,  $\{\ob E, \om E^2, 1\}$.

In the rest of this section, we show that each VEV in Eqs.~(\ref{eq:flavvevs}) can be expressed as a unique and simultaneous invariant eigenstate of a set of group elements of $\sgm\times\bt\times\bt$. These elements constitute a subgroup of the flavor group. In other words, each VEV is uniquely identified by a specific subgroup of the flavor group.

Let us study the VEV of the flavon $\Delta$. Consider the group element
\begin{equation}\label{eq:opd1}
\mathcal {O}_{C\bind}= \cro \crt \times \cro\!^2\, \crt\!^2\, \crf\!^2 
\end{equation}
acting on $\langle\bind\rangle$, Eq.~(\ref{eq:flavvevs}). The direct product, Eq.~(\ref{eq:opd1}) (corresponding to $\bt\times\bt$), acts on the two indices of  $\langle\bind\rangle$. As a matrix equation, the operation of this group element on the VEV can be written as
\begin{equation}\label{eq:opd1mat}
\left(\begin{matrix}
	1 & 0 & 0 & 0 & 0 & 0\\
     	0 & \ob & 0 & 0 & 0 & 0\\
      	0 & 0 & \om & 0 & 0 & 0\\
	0 & 0 & 0 & \om & 0 & 0\\
	0 & 0 & 0 & 0 & \ob & 0\\
	0 & 0 & 0 & 0 & 0 & 1
\end{matrix}\right)\left(\begin{matrix}
	1 & 0 & 0 & 1 & 0 & 0\\
     	0 & 1 & 0 & 0 & 1 & 0\\
      	0 & 0 & 1 & 0 & 0 & 1\\
	0 & 0 & 1 & 0 & 0 & 1\\
	0 & 1 & 0 & 0 & 1 & 0\\
	1 & 0 & 0 & 1 & 0 & 0
\end{matrix}\right)\left(\begin{matrix}
	1 & 0 & 0 & 0 & 0 & 0\\
     	0 & \om & 0 & 0 & 0 & 0\\
      	0 & 0 & \ob & 0 & 0 & 0\\
	0 & 0 & 0 & 1 & 0 & 0\\
	0 & 0 & 0 & 0 & \om & 0\\
	0 & 0 & 0 & 0 & 0 & \ob
\end{matrix}\right)^T
\end{equation}
where the rows and columns of $\langle\bind\rangle$ correspond to the first and the second $\bt$ in the direct product. It is clear that this operation multiplies all the vanishing elements in the VEVs with $\om$ or $\ob$. Therefore, invariance of the VEV under $\mathcal {O}_{C\bind}$ ensures that these elements vanish.

Consider the group element
\begin{equation}\label{eq:opd2}
\mathcal {O}_{Dc\bind}= I \times  \Dt
\end{equation}
where $I$ is the identity. As a matrix equation, the operation of this element on the VEV can be written as
\begin{equation}\label{eq:opd2mat}
\left(\begin{matrix}
	1 & 0 & 0 & 0 & 0 & 0\\
     	0 & 1 & 0 & 0 & 0 & 0\\
      	0 & 0 & 1 & 0 & 0 & 0\\
	0 & 0 & 0 & 1 & 0 & 0\\
	0 & 0 & 0 & 0 & 1 & 0\\
	0 & 0 & 0 & 0 & 0 & 1
\end{matrix}\right)\left(\begin{matrix}
	1 & 0 & 0 & 1 & 0 & 0\\
     	0 & 1 & 0 & 0 & 1 & 0\\
      	0 & 0 & 1 & 0 & 0 & 1\\
	0 & 0 & 1 & 0 & 0 & 1\\
	0 & 1 & 0 & 0 & 1 & 0\\
	1 & 0 & 0 & 1 & 0 & 0
\end{matrix}\right)\left(\begin{matrix}
	0 & 0 & 0 & 1 & 0 & 0\\
     	0 & 0 & 0 & 0 & 1 & 0\\
      	0 & 0 & 0 & 0 & 0 & 1\\
	1 & 0 & 0 & 0 & 0 & 0\\
	0 & 1 & 0 & 0 & 0 & 0\\
	0 & 0 & 1 & 0 & 0 & 0
\end{matrix}\right)^T.
\end{equation}
This operation interchanges the columns $1$, $2$, $3$ of the VEV with the columns $4$, $5$, $6$, respectively. Invariance under this operation ensures that the columns $1$, $2$, $3$ become equal to the columns $4$, $5$, $6$, respectively. This condition is satisfied by our VEV.

Now consider the group element
\begin{equation}\label{eq:opd3}
\mathcal {O}_{Dr\bind}= \Do \times  I.
\end{equation}
As a matrix equation, its operation on the VEV can be written as
\begin{equation}\label{eq:opd3mat}
\left(\begin{matrix}
	0 & 0 & 0 & 0 & 0 & 1\\
     	0 & 0 & 0 & 0 & 1 & 0\\
      	0 & 0 & 0 & 1 & 0 & 0\\
	0 & 0 & 1 & 0 & 0 & 0\\
	0 & 1 & 0 & 0 & 0 & 0\\
	1 & 0 & 0 & 0 & 0 & 0
\end{matrix}\right)\left(\begin{matrix}
	1 & 0 & 0 & 1 & 0 & 0\\
     	0 & 1 & 0 & 0 & 1 & 0\\
      	0 & 0 & 1 & 0 & 0 & 1\\
	0 & 0 & 1 & 0 & 0 & 1\\
	0 & 1 & 0 & 0 & 1 & 0\\
	1 & 0 & 0 & 1 & 0 & 0
\end{matrix}\right)\left(\begin{matrix}
	1 & 0 & 0 & 0 & 0 & 0\\
     	0 & 1 & 0 & 0 & 0 & 0\\
      	0 & 0 & 1 & 0 & 0 & 0\\
	0 & 0 & 0 & 1 & 0 & 0\\
	0 & 0 & 0 & 0 & 1 & 0\\
	0 & 0 & 0 & 0 & 0 & 1
\end{matrix}\right)^T.
\end{equation}
This operation interchanges the rows $1$, $2$, $3$ of the VEV with the rows $6$, $5$, $4$, respectively. Invariance under this operation ensures that the rows $1$, $2$, $3$ become equal to the rows $6$, $5$, $4$, respectively. This condition is also satisfied by our VEV.

Finally we consider the group element
\begin{equation}\label{eq:opd4}
\mathcal {O}_{E\bind}= \Et \times  \Eo.
\end{equation}
As a matrix equation, its operation on the VEV is
\begin{equation}\label{eq:opd4mat}
\left(\begin{matrix}
	0 & 0 & 1 & 0 & 0 & 0\\
     	1 & 0 & 0 & 0 & 0 & 0\\
      	0 & 1 & 0 & 0 & 0 & 0\\
	0 & 0 & 0 & 0 & 1 & 0\\
	0 & 0 & 0 & 0 & 0 & 1\\
	0 & 0 & 0 & 1 & 0 & 0
\end{matrix}\right)\left(\begin{matrix}
	1 & 0 & 0 & 1 & 0 & 0\\
     	0 & 1 & 0 & 0 & 1 & 0\\
      	0 & 0 & 1 & 0 & 0 & 1\\
	0 & 0 & 1 & 0 & 0 & 1\\
	0 & 1 & 0 & 0 & 1 & 0\\
	1 & 0 & 0 & 1 & 0 & 0
\end{matrix}\right)\left(\begin{matrix}
	0 & 0 & 1 & 0 & 0 & 0\\
     	1 & 0 & 0 & 0 & 0 & 0\\
      	0 & 1 & 0 & 0 & 0 & 0\\
	0 & 0 & 0 & 0 & 0 & 1\\
	0 & 0 & 0 & 1 & 0 & 0\\
	0 & 0 & 0 & 0 & 1 & 0
\end{matrix}\right)^T.
\end{equation}
This operation cycles various sets of three elements of the VEV. There are 12 such sets. These include $(11,22,33)$, $(14,25,36)$, $(61,52,43)$, $(64,55,46)$ where the pairs denote the indices of $\langle\bind\rangle$.  Invariance under this operation ensures that the elements within a set are equal to one another. Our VEV satisfies this condition also.  

The flavon, $\bind$, transforms in a vector space on which the flavor group acts.\footnote{The left and the right multiplying matrices acting on $\bind$ form an unfaithful matrix representation of the full flavour group in this space.} In a vector space, a state that does not change under the operation of a group element is an invariant eigenstate of the element. $\langle\bind\rangle$, Eq.~(\ref{eq:flavvevs}), is an invariant eigenstate of four group elements, Eqs.~(\ref{eq:opd1}), (\ref{eq:opd2}), (\ref{eq:opd3}), and (\ref{eq:opd4}). Invoking the condition that the VEV is an invariant eigenstate of all these four elements fixes the orientation of the VEV in the vector space ensuring that the VEV is proportional to $\langle\bind\rangle$, Eqs.~(\ref{eq:flavvevs}).

The group elements $\mathcal {O}_{C\bind}$ and $\mathcal {O}_{E\bind}$, Eqs.~(\ref{eq:opd1}) and (\ref{eq:opd4}), generate two $C_3$ groups. This is evident by inspecting the corresponding matrix operations in Eqs.~(\ref{eq:opd1mat}) and (\ref{eq:opd4mat}). Similarly the group elements $\mathcal {O}_{Dc\bind}$ and $\mathcal {O}_{Dr\bind}$, Eqs.~(\ref{eq:opd2}) and (\ref{eq:opd3}), generate two $C_2$ groups, as is clear from the corresponding matrix operations, Eqs.~(\ref{eq:opd2mat}) and (\ref{eq:opd3mat}).  To denote the action of a direct product element, we used left and right multiplications with the corresponding matrices. To represent a direct product element using a single matrix, we need to obtain the Kronecker product of the left and the right matrices. It can be shown that the four Kronecker product matrices, corresponding to the four direct product elements, Eqs.~(\ref{eq:opd1}), (\ref{eq:opd2}), (\ref{eq:opd3}, and (\ref{eq:opd4}), commute with each other\footnote{Taken separately, the left (the right) parts of the direct product elements, i.e.,~the left (the right) multiplying matrices, do not commute with each other.}; i.e., they generate the subgroup $C_3\times C_2 \times C_2 \times C_3$. Therefore, the flavon VEV $\langle\bind\rangle$ breaks $\bt\times\bt$ into $C_3\times C_2 \times C_3 \times C_2= C_6 \times C_6$. To summerize, the $C_6 \times C_6$ subgroup generated by $\mathcal {O}_{C\bind}$, $\mathcal {O}_{Dc\bind}$, $\mathcal {O}_{Dr\bind}$, and $\mathcal {O}_{E\bind}$ remains as the residual symmetry of the VEV, and it uniquely defines the VEV (up to multiplication by an overall complex constant). In the language of orbits and stabilizers, the $C_6 \times C_6$ subgroup is the stabilizer of $\langle\bind\rangle$ and it uniquely defines the orbit of $\langle\bind\rangle$.

Now we turn our attention to the flavons, $\phia$ and $\phib$. Consider the group element
\begin{equation}\label{eq:opp1}
\mathcal {O}_{C\phi}= \cro \crt \crf \times C
\end{equation}
operating on the flavon VEVs $\langle\phia\rangle$ or $\langle\phib\rangle$, Eqs.~(\ref{eq:flavvevs}). Note that $C$ is an element of $\sgm$, Eqs.~(\ref{eq:gen3}). The $\bt$ and $\sgm$ parts of $\mathcal {O}_{C\phi}$ act on the corresponding indices of the flavon. As a matrix equation, the operation of $\mathcal {O}_{C\phi}$ on the VEV can be written as
\begin{equation}\label{eq:opp1mat}
\left(\begin{matrix}
	1 & 0 & 0 & 0 & 0 & 0\\
     	0 & \ob & 0 & 0 & 0 & 0\\
      	0 & 0 & \om & 0 & 0 & 0\\
	0 & 0 & 0 & 1 & 0 & 0\\
	0 & 0 & 0 & 0 & \ob & 0\\
	0 & 0 & 0 & 0 & 0 & \om
\end{matrix}\right)\left(\begin{matrix}
	\Ta & 0 & 0\\
     	0 & \om & 0\\
      	0 & 0 & \Tb\\
	-i & 0 & 0\\
	0 & \ob & 0\\
	0 & 0 & i
\end{matrix}\right)\left(\begin{matrix}1 & 0 & 0\\
       0 & \om & 0\\
       0 & 0 & \ob
\end{matrix}\right)^T.
\end{equation}
It is clear that this operation multiplies all the vanishing elements in the VEV with $\om$ or $\ob$. Therefore, invariance of the VEV under $\mathcal {O}_{C\phi}$ ensures that these elements vanish.

Consider the group element
\begin{equation}\label{eq:opp2}
\mathcal {O}_{D\phi}= \cro \crt\!^2 \crr \crf\!^2 \crv \cer\!^3 \cef\!^3 \Dt \times I.
\end{equation}
As a matrix equation, its operation on the VEV is
\begin{equation}\label{eq:opp2mat}
\left(\begin{matrix}
	0 & 0 & 0 & -\Tb & 0 & 0\\
     	0 & 0 & 0 & 0 & \ob & 0\\
      	0 & 0 & 0 & 0 & 0 & -\Ta\\
	-\Ta & 0 & 0 & 0 & 0 & 0\\
	0 & \om & 0 & 0 & 0 & 0\\
	0 & 0 & -\Tb & 0 & 0 & 0
\end{matrix}\right)\left(\begin{matrix}
	\Ta & 0 & 0\\
     	0 & \om & 0\\
      	0 & 0 & \Tb\\
	-i & 0 & 0\\
	0 & \ob & 0\\
	0 & 0 & i
\end{matrix}\right)\left(\begin{matrix}1 & 0 & 0\\
       0 & 1 & 0\\
       0 & 0 & 1
\end{matrix}\right)^T.
\end{equation}
This operation interchanges the rows, $1$, $2$, $3$, of the VEV with the rows, $4$, $5$, $6$, respectively, along with multiplication of these rows with certain specific values of phases. Invariance under this operation ensures that the elements in the upper and the corresponding lower rows in the VEV have the same magnitude, but differ by specific phases. Our VEV satisfies this condition.   

Finally consider the group element
\begin{equation}\label{eq:opp3}
\mathcal {O}_{E\phi}= \cro\!^2 \crt \crr\!^2 \ceo\!^7 \cet\!^2 \cer \cef\!^4 \Eo\!^2 \times E
\end{equation}
As a matrix equation, its operation on the VEV is
\begin{equation}\label{eq:opp3mat}
\left(\begin{matrix}
	0 & \ob \Ta & 0 & 0 & 0 & 0\\
     	0 & 0 & \om \Ta & 0 & 0 & 0\\
      	-i & 0 & 0 & 0 & 0 & 0\\
	0 & 0 & 0 & 0 & -i \om & 0\\
	0 & 0 & 0 & 0 & 0 & -i \ob\\
	0 & 0 & 0 & -1 & 0 & 0
\end{matrix}\right)\left(\begin{matrix}
	\Ta & 0 & 0\\
     	0 & \om & 0\\
      	0 & 0 & \Tb\\
	-i & 0 & 0\\
	0 & \ob & 0\\
	0 & 0 & i
\end{matrix}\right)\left(\begin{matrix}0 & 1 & 0\\
       0 & 0 & 1\\
       1 & 0 & 0
\end{matrix}\right)^T
\end{equation}
This operation cycles various sets of three elements of the VEV, along with multiplying these elements with specific phases. There are six such sets in the VEV. These include $(11,22,33)$ and $(41,52,63)$, where the pairs denote the indices of the VEV.  Invariance under this operation ensures that the elements within a set are equal to one another in magnitude, but differ by specific phases. Our VEV satisfies this condition also.  

Invoking the condition that the VEVs of the flavons $\phia$ and $\phib$, Eq.~(\ref{eq:flavvevs}), are invariant under $\mathcal {O}_{C\phi}$, $\mathcal {O}_{D\phi}$, and $\mathcal {O}_{E\phi}$ uniquely defines the orientation of the VEVs in the flavor space. The group elements $\mathcal {O}_{C\phi}$, $\mathcal {O}_{D\phi}$, and $\mathcal {O}_{E\phi}$, Eqs.~(\ref{eq:opp1}), (\ref{eq:opp2}), and (\ref{eq:opp3}), generate $C_3$, $C_2$, and $C_3$ groups, respectively. This is evident by inspecting the corresponding matrix operations in Eqs.~(\ref{eq:opp1mat}), (\ref{eq:opp2mat}), and (\ref{eq:opp3mat}). These three elements also commute with each other, so that they generate the subgroup $C_3 \times C_2 \times C_3 = C_6 \times C_3$. To prove that they commute we need to calculate the Kronecker product matrices, as we discussed in the case of the $\bind$ flavon. To summarize, the $C_6 \times C_3$ subgroup generated by $\mathcal {O}_{C\phi}$, $\mathcal {O}_{D\phi}$, and $\mathcal {O}_{E\phi}$ remains as the residual symmetry of $\langle\phia\rangle$ and $\langle\phib\rangle$ after SSB, and it uniquely defines these VEVs (up to multiplication by an overall complex constant); i.e.,~the VEVs and their orbits are uniquely defined in terms of the $C_6 \times C_3$ stabilizer groups.

Since the neutrinos, $\nu_R$, form a triplet under $\sgm$, the individual states, $\nu_{R1}$, $\nu_{R2}$ and $\nu_{R3}$ correspond to the flavor basis states, $(1,0,0)^T$, $(0,1,0)^T$, and $(0,0,1)^T$ respectively. These states are the invariant eigenstates of the group elements $C$, $E^2CE$, and $ECE^2$, respectively where the group generators are given in Eqs.~(\ref{eq:gen3}). Individually, the above-mentioned group elements form $C_3$ subgroups of $\sgm$. To summarize, we have shown that the flavon VEVs as well as the neutrino states can be uniquely defined in terms of their symmetry properties. They are expressed as the invariant eigenstates of specific group elements that form specific subgroups of the flavor symmetry group. Thus the flavor structure of our model is entirely determined by the subgroup structure of the flavor symmetry group. It should be noted that, even though we have used matrix representations in convenient bases, our formalism is manifestly basis independent, i.e. expressible in terms of the abstract group generators.

A motivation for using discrete groups in model building is that they allow us to choose the flavon VEVs from among a discrete set of alignments determined by their symmetries. It is interesting to analyze how probable it is to obtain the given VEVs from the set of minima of the flavon potential through SSB. In this paper, we did not analyze the flavon potential for our model. Nevertheless, let us assume that a potential can be constructed that leads to the required VEVs, Eqs.~(\ref{eq:vevmutau}) and (\ref{eq:flavvevs}), through SSB. Let us also assume that the different irreducible multiplets ($\phi_\mu$, $\phi_\tau$, $\phia$, $\phib$, $\bind$) are decoupled (at least at the renormalizable level) in the potential so that the minima of the potential correspond to the VEVs of the orbits of each multiplet independently of the others. 

At the beginning of this section, we showed that the VEV, $\langle \phi_\mu \rangle$, has a $C_3$ residual symmetry. The flavon $\phi_\mu$ transforms under $\sgm$ and also under the $C_3$ group given in Table~\ref{tab:flavourcontent}. Therefore, a total of $216\times 3$ group elements act on $\phi_\mu$. Since the VEV has a $C_3$ residual symmetry, the orbit of $\langle \phi_\mu \rangle$ will have $(216\times 3)/3$ distinct points. Therefore, the probability of $\phi_\mu$ acquiring the given VEV through SSB is 1 in 216. It should also be noted that the alignments $\langle \phi_\mu \rangle$, $\om \langle \phi_\mu \rangle$, and $\ob \langle \phi_\mu \rangle$ (three different minima in the orbit) leads to the same phenomenology. As a result, the probability effectively increases to $1/72$. Similarly, we have a $1/72$ probability to obtain $\langle \phi_\tau \rangle$, $\om \langle \phi_\tau \rangle$, or $\ob \langle \phi_\tau \rangle$.

The flavon $\phia$ transforms under both $\sgm$ and $\bt$. We have found that $\bt$ has $N=8^43^52^23^2$ elements. Hence a total of $216\times N$ group elements act on $\phia$. we have also found that $\langle \phia \rangle$ has a $C_6\times C_3$ residual symmetry. Therefore, the orbit of $\langle \phia \rangle$ will have $216\times N/18 = 12N$ distinct points. Similarly, there will also be $12N$ points in the orbit of $\langle \phib \rangle$. The flavon $\bind$ transforms under two copies of $\bt$,\footnote{$\bind$ also transforms under a $C_2$, Table~\ref{tab:flavourcontent}. We have not included this transformation in our analysis because both $\langle\bind\rangle$ and $-\langle\bind\rangle$ lead to the same phenomenology and hence the effective probability remains unaffected.} and its VEV has a $C_6\times C_6$ residual symmetry. As a result, there will be $N^2/36$ points in the orbit of $\langle\bind\rangle$. Taking the orbits of $\langle\phia\rangle$, $\langle\phib\rangle$, and $\langle\bind\rangle$ together, we obtain $12N\times 12N\times N^2/36 = 4N^4$ distinct points. So we conclude that the probability of obtaining the VEVs, Eqs~(\ref{eq:flavvevs}), is 1 in $4N^4$. Consider a transformation $\langle \phia \rangle \rightarrow g_1 \langle \phia \rangle$, $\langle \phib \rangle \rightarrow g_2 \langle \phia \rangle$, and $\langle \bind \rangle \rightarrow (g_1^T)^{-1} \langle \phia \rangle g_2^{-1}$ where $g_1$ and $g_2$ are elements of the two copies of $\bt$, respectively. The Majorana mass matrix, Eq.~(\ref{eq:majexpr}), and hence the phenomenology remain unaffected by this transformation. There are a total of $N^2$ elements in this transformation. Therefore the effective probability increases to $1/(4N^2)$.

Considering all the flavons ($\phi_\mu, \phi_\tau, \phia, \phib, \bind$), we have an effective probability of $(1/72)^2\times1/(4N^2)$ to achieve the phenomenology described in our model. This probability is very small. However, it is still better than the model described in Ref.~\cite{Krishnan:2018tja} where the same phenomenology is obtained by tuning the parameters in the potential so as to get the required vacuum alignment. As argued earlier in this paper, such a tuning corresponds to a vanishing effective probability.

\section{The new framework}
\label{sec:aux}

We have introduced a new framework in flavor physics which involves the flavor group, $G_f$, expressed as a direct product, $G_f=G_r\times G_x$, where $G_x$, which we call the auxiliary group, is a suitable discrete group under which only the scalars (flavons) transform nontrivially. The scalars may also transform under $G_r$.  In contrast, the fermions transform nontrivially only under $G_r$. In this paper, we have $G_r=\sgm\times C_3\times C_4$ and $G_x = \bt\times \bt$. But in general, $G_r$ can be any discrete group commonly studied in the literature such as $A_4$, $S_4$, $A_5$, $\Delta(3n^2)$, and $\Delta(6n^2)$~\cite{Petcov:2018snn, Pramanick:2017fdq, Pramanick:2017wry, Borah:2018nvu, Borah:2017qdu, King:2018fke, Sruthilaya:2017mzt, Karmakar:2016cvb, Ky:2016rzl, Zhang:2015vle, deAnda:2017yeb, deAnda:2018oik, Mukherjee:2017pzq, King:2016yvg, Vien:2016jkz, Shimizu:2015tta, Feruglio:2013hia, Krishnan:2012me, Ding:2017hdv, Yao:2016zev, Joshipura:2016hvn, Girardi:2015rwa, Joshipura:2015dsa, Turner:2015uta, DiIura:2015kfa, Li:2015jxa, Lam:2014kga, CarcamoHernandez:2018iel, CarcamoHernandez:2017owh, Bjorkeroth:2015uou, deMedeirosVarzielas:2017ote, Escobar:2008vc, Ishimori:2014nxa, Harrison:2014jqa, King:2013vna, Vien:2016tmh, Varzielas:2016zjc, Rodejohann:2017lre, Liu:2013oxa, Dicus:2010iq}. With three families of fermions, $G_r$ should be a discrete subgroup of $U(3)$ (together with several Abelian discrete groups, i.e.,~$C_n$'s). On the other hand, $G_x$ can be any discrete group with the requisite symmetries, but not necessarily a $U(3)$ subgroup. Until now, only the discrete subgroups of $U(3)$ have been used as flavor symmetry groups in the literature. By introducing the concept of the auxiliary group, which is no longer required to be a $U(3)$ subgroup, we have considerably broadened the model builders' toolkit. 

In this framework, the first step is to assume a flavon, which is an irreducible multiplet under $G_r\times G_x$. Then we identify a specific flavon alignment that remains invariant under the group transformations involving both $G_r$ and $G_x$. These transformations are used to fully define the alignment. Two or more flavons under $G_r\times G_x$ with similar uniquely defined alignments are also assumed. These flavons are coupled together to obtain an object that is an irreducible multiplet under $G_r$, but an invariant singlet under $G_x$. The alignments of the constituent flavons give rise to a unique alignment of this object. It should be noted that this unique alignment of the effective $G_r$ multiplet may not possess any residual symmetry under $G_r$ itself even though the alignments of the constituent flavons are fully defined in terms of their residual symmetries under $G_r\times G_x$. 

For a further demonstration of this procedure refer to Appendix~B, where we construct a VEV for the triplet ($\rtp$) of $S_4$ by using the dihedral group as the auxiliary group, i.e. $G_r=S_4$ and $G_x = D_{2n}$. We obtain a VEV of the form $\langle\rtp\rangle\propto\left( \cos (m\theta) ,\cos  (m\theta+\frac{2\pi}{3}) ,\cos (m\theta-\frac{2\pi}{3}) \right)$ where $\theta=\frac{2\pi}{n}$. This shows that even the humble dihedral group can produce an alignment for the $S_4$ triplet that has not been considered in the literature. We hope that this framework will stimulate research involving various choices of auxiliary groups combined with the commonly studied subgroups of $U(3)$. This may lead to novel choices of vacuum alignments for the irreducible multiplets of the $U(3)$ subgroups and new textures of mass matrices.

An important feature of this framework is that the flavons transform under a larger group ($G_r\times G_x$) compared to the fermions ($G_r$). Certain subgroups of $G_r\times G_x$ remain unbroken by the individual flavon VEVs. In some cases, the same residual symmetry may be present in the VEVs of all the flavons in the model; i.e.,~this symmetry is preserved for the whole Lagrangian. If the residual symmetry is a subgroup of the auxiliary group alone, then the fermions themselves will not transform under it. These conditions make the flavons stable against decay to fermions (Standard Model particles). As a result, the lightest flavon can become a dark-matter candidate. We will investigate such a model in a future work. 

\section{Summary}
\label{sec:summary}

In an earlier publication, we showed that a fully constrained Majorana mass matrix can be constructed using a sextet of $\sgm$. Specific VEVs for this sextet led to $\tpm$ mixing with $\phi=\pm\pi/16$ and the neutrino mass ratios given in Eq.~(\ref{eq:numass}). In this paper, we obtain these VEVs entirely using the principles of symmetries. To achieve this, we propose a new discrete symmetry group, $\bt$. Flavons $\phia$, $\phib$, and $\bind$ which transform under the expanded flavor group $\sgm\times\bt\times\bt$ are introduced. The VEV of each of these flavons is uniquely identified as the invariant eigenstate of several elements of the expanded flavor group. Such a set of elements generates a specific subgroup of $\sgm\times\bt\times\bt$ which forms the residual symmetry of the given VEV. The VEVs of $\phia$, $\phib$, and $\bind$ are coupled together to obtain the VEV of the sextet of $\sgm$. By imposing the condition that the VEVs of the constituent flavons are invariant eigenstates under the simultaneous action of $\sgm$ and $\bt$, we make the VEV of the sextet of $\sgm$ implicitly dependent on $\bt$.  

Using the irreducible triplet of $S_4$ group as an example, we show that flavon alignments fully defined by the residual symmetries of $S_4$ form unique orbits. On the other hand, alignments obtained by extremizing flavon potentials may not always be unique; i.e.,~by carefully adjusting the parameters in the potential we may obtain almost any arbitrary vacuum alignment having no residual symmetries. We argue that constructing such arbitrary  potentials goes against the spirit of using discrete symmetries to explain the flavor structure. Yet, to obtain phenomenologically viable models, we may have to resort to using a VEV (of an irreducible flavon multiplet) having no apparent residual symmetry. It is in this context that we introduce the new framework in which the flavor group is obtained as the direct product, $G_f=G_r\times G_x$, where the flavons transform under both $G_r$ and $G_x$ while the fermions transform only under $G_r$. By coupling together several flavons that transform under $G_r\times G_x$, we obtain an effective irreducible multiplet that transforms only under $G_r$. We define the alignments of the constituent flavons in terms of  the residual symmetries under $G_r\times G_x$. As a result we uniquely obtain the alignment of the effective $G_r$ multiplet as well, even though this multiplet may not possess any residual symmetry under $G_r$. We hope that more models are constructed in this framework leading to interesting predictions in flavor physics.

\section*{Acknowledgments}

I thank Paul Harrison and Bill Scott for the helpful discussions and Aidan Wiederhold for helping with making the plot. I acknowledge the support from the University of Warwick and the hospitality of the Particle Physics Department at the Rutherford Appleton Laboratory. I thank the management of the School of the Good Shepherd, Thiruvananthapuram, for providing a convenient and flexible working arrangement conducive to research.

\section*{Appendix A: $S_4$ VEVs obtained from potentials} 

Here we use the group $S_4$ to construct a couple of toy models for mass matrices. We investigate the flavon potentials and show that they may or may not lead to VEVs with specific symmetry properties. In Sec.~II, we discussed the properties of the triplet representations, $\boldsymbol{3}$ and $\rtp$. These representations are faithful consisting of 24 distinct group elements. However, the doublet representation ($\boldsymbol{2}$) is not faithful. The $2\times 2$ matrices, Eqs.~(\ref{eq:s4dblt}), generate the dihedral group $D_6$ which forms a quotient group of $S_4$. $D_6$ represents the rotation as well as the reflection symmetries of an equilateral triangle as shown in Fig.~\ref{fig:triangles}. 

\begin{figure}[h]
\begin{center}
\includegraphics[trim={5cm 1cm 5cm 1cm},scale=0.32]{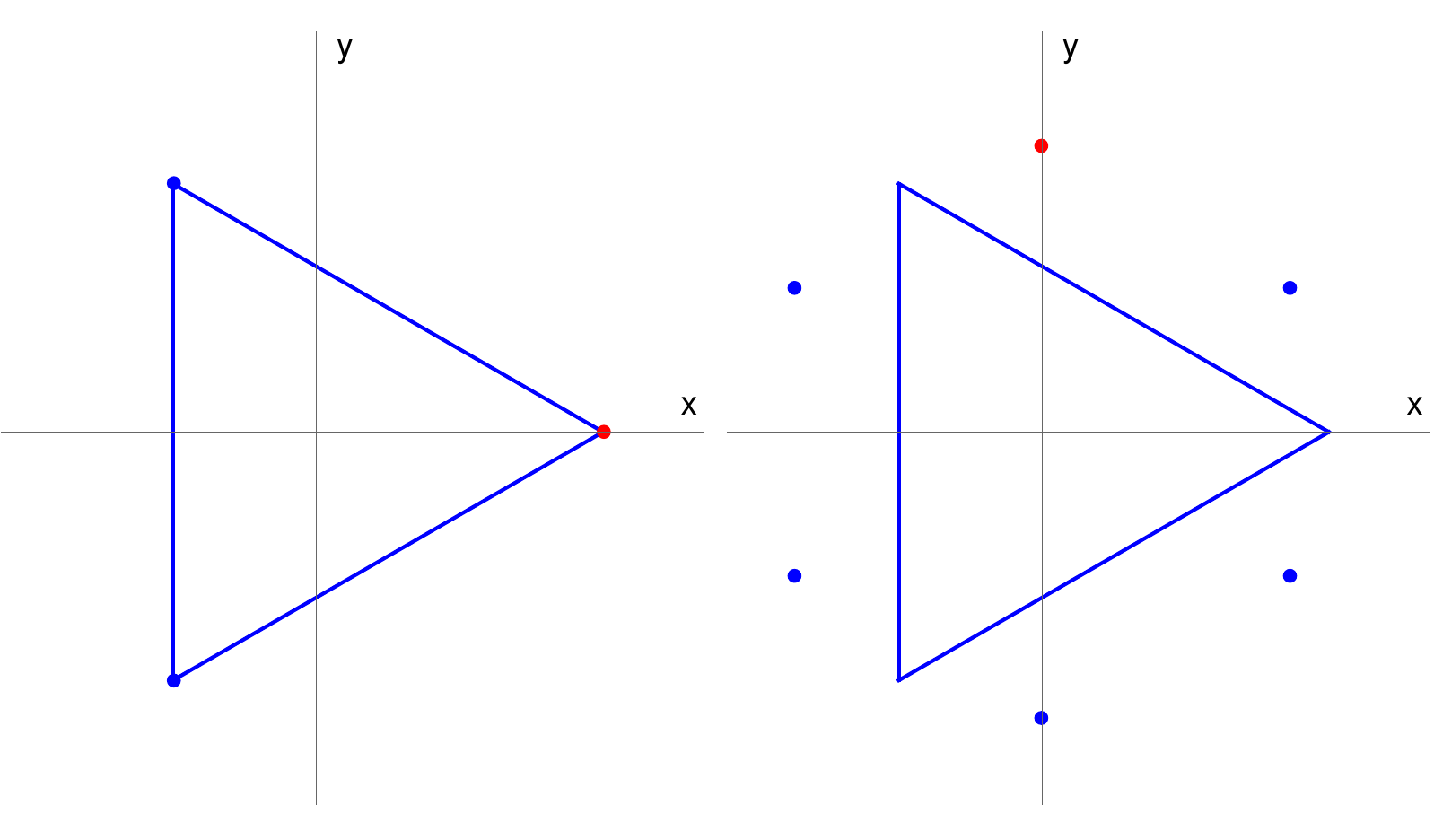}
\caption[The triangles]{The dots on the left and the right figures represent extrema of two cases of flavon potentials which have $D_6$ symmetry. In the left figure, the extrema are on vertices of the triangle. The VEV, denoted by the red dot, is aligned along the direction representing one of the reflection symmetries of the triangle. In the right figure, $D_6$ symmetry of the potential results in $6$ extremum points positioned symmetrically around the triangle. However, the VEV (the red dot) is not aligned along any special direction in relation to the triangle.}
\label{fig:triangles}
\end{center}
\end{figure} 

Let us define singlet ($\phi_s$), doublet ($\phi_d$), and triplet ($\phi_t$) flavons which transform as $\boldsymbol{1}$ (invariant), $\boldsymbol{2}$, and $\rtp$ respectively. They couple with the neutrino multiplets, Eqs.~(\ref{eq:numul1})--(\ref{eq:numul3}), to produce the $S_4$ invariant mass term,
\begin{equation}\label{eq:fullterm}
k_s (\nu_R\cdot\nu_R)_s \phi_s +k_d (\nu_R\cdot\nu_R)_d^T \phi_d+ k_t (\nu_R\cdot\nu_R)_t^T \phi_t,
\end{equation}
where $k_s$, $k_d$, and $k_t$ are the coupling constants. The flavons and the coupling constants in Eq.~(\ref{eq:fullterm}) can be written in a matrix form,
{\footnotesize
\begin{equation}\label{eq:fullmm}
k_s \phi_s I +
\left(\begin{matrix}k_d (2 \phi_{d1})&  k_t \phi_{t3} &  k_t \phi_{t2}\\
         k_t \phi_{t3} & k_d (-\phi_{d1}+\sqrt{3}\phi_{d2})&  k_t \phi_{t1}\\
         k_t \phi_{t2} &  k_t \phi_{t1} &  k_d (-\phi_{d1}-\sqrt{3}\phi_{d2})
\end{matrix}\right),
\end{equation}
}where we have expressed the doublet and the triplet flavons in terms of their components, i.e.,~$\phi_d=(\phi_{d1},\phi_{d2})^T$ and $\phi_t=(\phi_{t1},\phi_{t2},\phi_{t3})^T$. Substituting a specific vacuum alignment for the flavons in Eq.~(\ref{eq:fullmm}) produces the mass matrix. In the rest of this appendix, two examples are provided where we minimize flavon potentials to obtain the VEVs and the corresponding mass matrices. In Example~1 the VEVs can also be defined in terms of their symmetries while in Example~2 they do not have such symmetry properties.

\subsection*{Example~1}

It is straightforward to write a potential for the invariant flavon $\phi_s$,
\begin{equation}
(\phi_s-1)^2.
\end{equation}
Extremizing this potential leads to the VEV 
\begin{equation}
 \langle\phi_s\rangle=1.
\end{equation}

In order to construct a potential for the doublet flavon, we first consider the tensor product of two doublets.  It can be shown that the tensor product leads to another doublet, 
\begin{equation}
(\phi_d\phi_d)_d=\left((\phi_{d1}^2-\phi_{d2}^2), -2 \phi_{d1} \phi_{d2}\right)^T.
\end{equation}
Now we construct the potential, 
\begin{equation}
\left|(\phi_d\phi_d)_d-\phi_d\right|^2,
\end{equation}
where the operator $|\,\,|^2$ represents $()^T()$. This potential has three minima: $\phi_d=(1,0)$, $(-\frac{1}{2},\frac{\sqrt{3}}{2})$, and $(-\frac{1}{2},-\frac{\sqrt{3}}{2})$. They form the vertices of the equilateral triangle as shown in Fig.~\ref{fig:triangles} (left). We assume that the flavon acquires one of these minima as its VEV,
\begin{equation}\label{eq:dbltvev1}
\langle\phi_d\rangle=(1,0)^T.
\end{equation}
This VEV breaks $D_6$ to one of its subgroups, $C_2$, generated by
\begin{equation}
PQ\equiv
\left(\begin{matrix}1 & 0\\
       0 & -1
\end{matrix}\right),
\end{equation}
where $E$ and $F$ are given in Eqs.~(\ref{eq:s4dblt}). $C_2$ represents the reflection symmetry of the triangle which keeps $(1,0)^T$ invariant. Conversely, the vacuum alignment, $(1,0)^T$, can be uniquely identified by this residual $C_2$ symmetry.

Now we construct a potential for the triplet flavon, $\phi_t$, which transforms as a $\rtp$. From the tensor product of two $\phi_t$ triplets, we obtain the second order triplet,
\begin{equation}
(\phi_t\phi_t)_t=(\phi_{t2}\phi_{t3}, \phi_{t3}\phi_{t1}, \phi_{t1}\phi_{t2})^T,
\end{equation}
similar to Eq.~(\ref{eq:numul3}). Using $\phi_t$ and $(\phi_t\phi_t)_t$, we construct the potential, 
\begin{equation}
\left(|\phi_t|^2-1\right)^2+k \left|(\phi_t\phi_t)_t\right|^2.
\end{equation}
This potential has six minima $\phi_t=(\pm1,0,0)$, $(0,\pm1,0)$ and $(0,0,\pm1)$. These are the face centers of the cube shown in Fig.~\ref{fig:cubes} (left). We assume that the flavon acquires one of these minima as its VEV,
\begin{equation}\label{eq:trpvev1}
\langle\phi_t\rangle=(1,0,0)^T.
\end{equation}
This VEV breaks $S_4$ to one of its subgroups, $C_2\times C_2$, generated by $(PQ)^2$ and $QPQ^2$, Eqs.~(\ref{eq:genc2c2}), as discussed in Sec.~II. Therefore, the VEV is uniquely identified by this residual $C_2\times C_2$ symmetry.

Substituting the VEVs, $\langle\phi_s\rangle$, $\langle\phi_d\rangle$, and $\langle\phi_t\rangle$ in Eq.~(\ref{eq:fullmm}), we obtain the mass matrix,
\begin{equation}\label{eq:unrealmm}
\left(\begin{matrix} k_s+2 k_d&  0 &  0\\
         0 & k_s -k_d&  k_t\\
         0 &  k_t & k_s -k_d
\end{matrix}\right).
\end{equation}
This matrix is diagonalized using the unitary matrix,
\begin{equation}
\left(\begin{matrix}1 & 0 & 0\\
       0 &\frac{1}{\sqrt{2}} & \frac{-1}{\sqrt{2}}\\
       0 & \frac{1}{\sqrt{2}} & \frac{1}{\sqrt{2}}
\end{matrix}\right),
\end{equation}
which provides a bimaximal contribution to mixing. By a suitable selection of the coupling constants, $k_s$, $k_d$, $k_t$, we can obtain any set of values for the masses without affecting the mixing part.

\subsection*{Example~2}

In this example we construct potentials for the flavons $\phi_d$ and $\phi_t$ leading to VEVs that leave no residual symmetries. For $\phi_d$, we use the potential
\begin{equation}\label{eq:dblflavpot}
(|\phi_d|^2-1)^2+\frac{1}{\Lambda^2}\left(\phi_d^T(\phi_d\phi_d)_d\right)^2,
\end{equation}
where the scale $\Lambda$ is added with the higher dimensional term. This potential has six minima which are of the form 
\begin{equation}\label{eq:dblmin}
\phi_d=g_i (0,1)^T,
\end{equation}
where $g_i$ are the six elements of the group $D_6$. These minima are shown in Fig.~\ref{fig:triangles} (right). We assume that flavon acquires one of the minima,
 \begin{equation}\label{eq:dblvev2}
\langle\phi_d\rangle=(0,1)^T.
\end{equation}
as its VEV. This VEV does not possess any residual symmetry of $D_6$.

For constructing the potential for the triplet flavon, we first obtain a doublet from the tensor product of two triplets. Similar to Eq.~(\ref{eq:numul2}), we obtain
\begin{equation}
(\phi_t\phi_t)_d=\left(2 \phi_{t1}^2-\phi_{t2}^2-\phi_{t3}^2,\sqrt{3}\phi_{t2}^2-\sqrt{3}\phi_{t3}^2 \right)^T.
\end{equation}
We construct the potential as
\begin{equation}\label{eq:tripflavpot}
\begin{split}
&\left(|\phi_t|^2-(1+\kappa_1^2+\kappa_2^2)\right)^2\\
&+\left|(\phi_t\phi_t)_d-\sqrt{3}(\kappa_1^2-\kappa_2^2)\phi_d+(2-\kappa_1^2-\kappa_2^2)(\phi_d\phi_d)_d    \right|^2,
\end{split}
\end{equation}
where $\kappa_1$ and $\kappa_2$ are arbitrary constants. In Eq.~(\ref{eq:tripflavpot}) we have coupled the doublet flavon, $\phi_d$, with the triplet flavon, $\phi_t$. Therefore, we should extremize Eq.~(\ref{eq:tripflavpot}), together with the potential for the doublet flavon, Eq.~(\ref{eq:dblflavpot}). If we substitute $\phi_d=(0,1)^T$ and $\phi_t=(1,\kappa_1,\kappa_2)^T$, both terms in Eq.~(\ref{eq:tripflavpot}) as well as in Eq.~(\ref{eq:dblflavpot}) vanish, indicating that these states of the flavons constitute a minimum of the potential. By transforming these flavon states under the action of $S_4$ we obtain further minima forming a discrete set. For $\phi_t$, these minima are shown in Fig.~\ref{fig:cubes} (right) where we have used $\kappa_1 = 0.50$ and $\kappa_2 = -0.85$. We select one of these minima,
 \begin{equation}\label{eq:trpvev2}
\langle\phi_t\rangle=(1,\kappa_1,\kappa_2)^T,
\end{equation}
as the VEV. As mentioned previously, this VEV does not possess any residual symmetry of $S_4$. Using the VEVs of the doublet and the triplet flavons we obtain the mass matrix,
\begin{equation}\label{eq:realmm}
\left(\begin{matrix} k_s&  k_t \kappa_2 &  k_t \kappa_1\\
         k_t \kappa_2 & k_s +\sqrt{3}k_d&  k_t\\
         k_t \kappa_1 &  k_t & k_s -\sqrt{3}k_d
\end{matrix}\right),
\end{equation}
which has more degrees of freedom compared to the previous case, Eq.~(\ref{eq:unrealmm}). By suitably tuning these free parameters, we can ensure that this mass matrix is consistent with the current neutrino masses and mixing data. However, we argue that since the VEVs, Eqs.~(\ref{eq:dblvev2}) and (\ref{eq:trpvev2}), have no apparent connection with the original flavor symmetry ($S_4$), we cannot claim that the texture of the resulting mass matrix has its origin in the aforementioned symmetry.

\section*{Appendix B: $S_4$ VEVs defined using the auxiliary group} 

In this section, we demonstrate the use of the auxiliary group to obtain unique vacuum alignments for the triplet ($\rtp$) of $S_4$. Here we use the dihedral group as the auxiliary group. The dihedral group, $D_{2n}$, is the symmetry group of a regular polygon with $n$ vertices. Such a polygon has $n$ rotational symmetries and $n$ reflection symmetries, so that the total number of elements of the group is $2n$. The generators for these rotations and reflections can be given by
\begin{equation}
R \equiv
\left(\begin{matrix}\cos \theta & \sin \theta\\
       -\sin \theta & \cos \theta
\end{matrix}\right), \quad F \equiv
\left(\begin{matrix}1 & 0\\
       0 & -1
\end{matrix}\right),
\end{equation}
where $\theta=\frac{2\pi}{n}$. The $n$ rotations and the $n$ reflections are given by
\begin{align}
R^m &\equiv
\left(\begin{matrix}\cos m\theta & \sin m\theta\\
       -\sin m\theta & \cos m\theta
\end{matrix}\right),\label{eq:d2ndblt1}\\
R^{-m} F R^m &\equiv
\left(\begin{matrix}\cos 2m\theta & \sin 2m\theta\\
       \sin 2m\theta & -\cos 2m\theta
\end{matrix}\right),\label{eq:d2ndblt2}
\end{align}
where $m=1...n$.  $R^m$ rotates by an angle $m\theta$ and $R^{-m} F R^m$ reflects about the axis $\left(\cos m\theta, \sin m\theta\right)^T$. Equations~(\ref{eq:d2ndblt1}) and (\ref{eq:d2ndblt2}) form the real doublet representation ($\boldsymbol{2}$) of $D_{2n}$.

For the purpose of using $D_{2n}$ as the auxiliary group with $S_4$, we assume that the angle $\theta$ divides $\frac{2\pi}{3}$; i.e.,~$n$ is a multiple of 3. This assumption implies that we have
\begin{equation}R_\omega= 
\left(\begin{matrix}\cos \frac{2\pi}{3} & \sin \frac{2\pi}{3}\\
       -\sin \frac{2\pi}{3} & \cos \frac{2\pi}{3}
\end{matrix}\right)
\end{equation}
as one of the group elements. $R_\omega$ and $F$ generate $D_6$ which forms a subgroup of the auxiliary group $D_{2n}$.

Our flavor group is $S_4\times D_{2n}$. We assume that there are two flavons, $\phia$ and $\phib$. They both transform as $\rtp$ under $S_4$ and $\boldsymbol{2}$ under $D_{2n}$. For these flavons, we use Latin and Greek indices to represent the $\rtp$ of $S_4$ and the $\boldsymbol{2}$ of $D_{2n}$, respectively, i.e.~$\phia_{\alpha i}$ and $\phib_{\alpha i}$. The two flavons are coupled together to obtain an object that transforms as a $\rtp$ under $S_4$ and an invariant under $D_{2n}$,
\begin{align}\label{eq:s4trip}
\begin{split}
(\phia\phib)_{\rtp}&=\sum_i (\phia_{2 i}\phib_{3 i}+\phia_{3 i}\phib_{2 i},\phia_{3 i}\phib_{1 i}+\phia_{1 i}\phib_{3 i},\\
&\,\,\quad\quad\quad\quad\quad\quad\quad\quad\quad\quad\quad\phia_{1 i}\phib_{2 i}+\phia_{2 i}\phib_{1 i}),
\end{split}
\end{align}
where the triplet ($\rtp$) is obtained as in Eq.~(\ref{eq:numul3}). Note that the doublet representation of $D_{2n}$ is real. Hence we construct the invariant by simply taking the scalar product (summation over $i$ in Eq.~(\ref{eq:s4trip})).

Consider the vacuum alignments,
\begin{align}
\langle \phia\rangle &=
\left(\begin{matrix}\cos 0 & \cos \frac{2\pi}{3} & \cos \frac{-2\pi}{3}\\
       \sin 0 & \sin \frac{2\pi}{3} & \sin \frac{-2\pi}{3}
\end{matrix}\right),\label{eq:phiphivevs1}\\
\langle \phib\rangle &=
\left(\begin{matrix}\cos m \theta & \cos (m \theta-\frac{2\pi}{3}) & \cos (m \theta+\frac{2\pi}{3})\\
       \sin m \theta & \sin (m \theta-\frac{2\pi}{3}) & \sin (m \theta+\frac{2\pi}{3})
\end{matrix}\right),\label{eq:phiphivevs2}
\end{align}
where the rows and the columns correspond to the Greek and the Latin indices, respectively. $\langle \phia\rangle$ remains invariant under the following transformations:
\begin{align}
R_\omega\langle \phia\rangle Q^2 &= \langle \phia\rangle,\label{eq:cubeorbits1}\\
F \langle \phia\rangle S & = \langle \phia\rangle,\label{eq:cubeorbits2}
\end{align}
where $S=PQ^2PQP$, Eq.~(\ref{eq:vertexstab}). The residual symmetry (the stabilizer subgroup) of $\langle \phia\rangle$ is generated by $Q^2\times R_\omega$ and $S\times F$ in the favor space of $S_4 \times D_{2n}$. It can be shown that these elements generate $D_6$. In other words, the $D_6$ subgroup of $S_4 \times D_{2n}$ generated by $Q^2\times R_\omega$ and $S\times F$ uniquely define $\langle \phia\rangle$.

$\langle\phia\rangle$ and $\langle\phib\rangle$, Eqs.~(\ref{eq:phiphivevs1}) and (\ref{eq:phiphivevs2}), are related by the group transformation,
\begin{equation}\label{eq:phiaphibrel}
R^m\langle \phib\rangle S = \langle\phia\rangle.
\end{equation}
Therefore, $\langle \phia\rangle$ and $\langle \phib\rangle$ share the same orbit. The residual symmetry of $\langle\phib\rangle$ can easily be found using Eqs.~(\ref{eq:cubeorbits1}) and (\ref{eq:cubeorbits2}), and (\ref{eq:phiaphibrel}), 
\begin{align}
R^{-m} R_\omega R^m\langle \phib\rangle SQ^2S^{-1} = R_\omega \langle \phib\rangle Q &= \langle \phib\rangle,\\
R^{-m} F R^m \langle \phib\rangle SSS^{-1} = R^{-m} F R^m \langle \phib\rangle S & = \langle \phib\rangle;
\end{align}
 i.e.,~$\langle \phib\rangle$ remains invariant under the group action of $Q\times R_\omega$ and $S\times R^{-m} F R^m$. These group elements generate another $D_6$ subgroup of $S_4 \times D_{2n}$, which uniquely defines $\langle \phib\rangle$ and forms its stabilizer. The orbit that includes $\langle \phia\rangle$ and $\langle \phib\rangle$ contains $\frac{24 \times 2n}{6}$ elements.

Now that $\langle \phia\rangle$ and $\langle \phib\rangle$ are uniquely defined using their residual symmetries, we move on to construct the $S_4$ triplet using Eq.~(\ref{eq:s4trip}). Substituting Eqs.~(\ref{eq:phiphivevs1}) and (\ref{eq:phiphivevs2}) in Eq.~(\ref{eq:s4trip}), we obtain
\begin{equation} \label{eq:dihedralalign}
(\langle \phia\rangle\langle \phib\rangle)_{\rtp}=2\left( \cos (m\theta) ,\cos  (m\theta+\frac{2\pi}{3}) ,\cos (m\theta-\frac{2\pi}{3}) \right).
\end{equation}
The above VEV ($\rtp$ of $S_4$) does not possess any residual symmetry of $S_4$ even though $\langle \phia\rangle$ and $\langle \phib\rangle$ are individually fully determined by their residual symmetries under $S_4\times D_{2n}$. $(\langle \phia\rangle\langle \phib\rangle)_{\rtp}$ produces an orbit with 24 elements in the space of $S_4$,  similar to Fig.~(\ref{fig:cubes}) (right). However, the flavon potential, Eq.~(\ref{eq:tripflavpot}), from which this figure is constructed, has arbitrary parameters $\kappa_1$ and $\kappa_2$.  By varying these parameters, we can construct infinitely many orbits and obtain the vacuum alignment along any direction we may choose. We argue that such arbitrary choices of alignments should be avoided. On the other hand, using $D_{2n}$ as the auxiliary group leads to alignments, Eq.~(\ref{eq:dihedralalign}), which are uniquely defined in terms of symmetries even if they fully break $S_4$. 

When the VEVs, Eqs.~(\ref{eq:vevtpmp}) and (\ref{eq:vevtpmm}), were proposed in Ref~\cite{Krishnan:2018tja}, we obtained them by extremizing a potential for the sextet flavon of $\sgm$. However, under $\sgm$ alone, these VEVs do not possess any residual symmetry. Their orbits have 216 elements with trivial stabilizers. In Ref.~\cite{Krishnan:2018tja}, these VEVs were obtained by simply adjusting the parameters in the potential. In this paper, we got rid of this arbitrariness by introducing the auxiliary group. 

\bibliography{DH12141.bib,noninspire.bib}

\end{document}